\documentclass[useAMS,usenatbib]{mn2e}

\usepackage{graphicx}
\usepackage{subfigure}
\usepackage{xcolor}
\usepackage{mathtext,bbm,amsmath,amsfonts,amssymb,indentfirst,syntonly,graphicx}
\usepackage{epstopdf}
\usepackage{mathtools}
\usepackage{slashbox}
\usepackage[english]{babel}
\usepackage{calc}
\usepackage{tikz}
\usepackage{color}
\usepackage{mathrsfs}
\usepackage{multirow}
\usepackage{bm}
\usepackage{caption}
\usepackage{natbib}
\usepackage{enumerate}
\usepackage{hyperref}

\title[Anisotropy of the Universe in Pantheon sample]{Anisotropy of the Universe via the Pantheon supernovae sample revisited}
\author[D. Zhao, Y. Zhou and Z. Chang]
        {Dong Zhao$^{1,2}$,
        Yong Zhou$^{1,2}$\thanks{Email: zhouyong@ihep.ac.cn} and
        Zhe Chang$^{1,2}$\\
$^{1}$Institute of High Energy Physics, Chinese Academy of Sciences, Beijing 100049, China\\
$^{2}$School of Physical Sciences, University of Chinese Academy of Sciences, Beijing 100049, China\\}
\begin{document}

\date{Accepted xxxx; Received xxxx; in original form xxxx}

\pagerange{\pageref{firstpage}--\pageref{lastpage}} \pubyear{2019}

\maketitle

\label{firstpage}

\begin{abstract}
We employ the hemisphere comparison (HC) method and the dipole fitting (DF) method to investigate the cosmic anisotropy in the recently released Pantheon sample of type Ia supernovae (SNe Ia) and five combinations among Pantheon. For the HC method, we find the maximum anisotropy level in the full Pantheon sample is $\mathrm{AL}_{max}=0.361\pm0.070$ and corresponding direction $(l,b)=({123.05^{\circ}}^{+11.25^{\circ}}_{-4.22^{\circ}}, {4.78^{\circ}}^{+1.80^{\circ}}_{-8.36^{\circ}})$. A robust check shows the statistical significance of maximum anisotropy level is about $2.1\sigma$. We also find that the Low-$z$ and SNLS subsamples have decisive impact on the overall anisotropy while other three subsamples have little impact. Moreover, the anisotropy level map significantly rely on the inhomogeneous distribution of SNe Ia in the sky. For the DF method, we find the dipole anisotropy in the Pantheon sample is very weak. The dipole magnitude is constrained to be less than $1.16\times10^{-3}$ at $95\%$ confidence level. However, the dipole direction is well inferred by MCMC method and it points towards $(l,b)=({306.00^{\circ}}^{+82.95^{\circ}}_{-125.01^{\circ}}, {-34.20^{\circ}}^{+16.82^{\circ}}_{-54.93^{\circ}})$. This direction is very close to the axial direction to the plane of SDSS subsample. It may imply that SDSS subsample is the decisive part to the dipole anisotropy in the full Pantheon sample. All these facts imply that the cosmic anisotropy found in Pantheon sample significantly rely on the inhomogeneous distribution of SNe Ia in the sky. More homogeneous distribution of SNe Ia is necessary to  search for a more convincing cosmic anisotropy.
\end{abstract}


\section{Introduction}\label{sec:introduction}\noindent
The cosmological principle assumes that the Universe is homogeneous and isotropic at large scales. Based on it, the $\Lambda$CDM model is well consistent with current cosmological observations. For instance, the cosmic microwave background (CMB) radiation observed by the Wilkinson Microwave Anisotropy Probe (WMAP) \citep{Bennett:2012zja,Hinshaw:2012aka} and Planck \citep{Ade:2015xua,Ade:2015lrj} satellites confirm the $\Lambda$CDM model with high precision. However, the cosmological principle has been challenged by a lot of accurate observations. An incomplete list includes the hemispherical power asymmetry of CMB temperature anisotropies \citep{Eriksen:2003db,Hansen:2004vq,Bennett:2012zja,Akrami:2014eta,Ade:2013nlj,Quartin:2014yaa}, the alignment of low-$\ell$ multipoles in the CMB temperature anisotropies \citep{Lineweaver:1996xa,Tegmark:2003ve,Bielewicz:2004en,Frommert:2009qw,Copi:2010na}, the parity asymmetry of low-$\ell$ multipoles in angular power spectrum of CMB temperature anisotropies \citep{Kim:2010gf,Kim:2010gd,Kim:2010st,Gruppuso:2010nd,Zhao:2013jya}, the large-scale alignment of quasar polarization vector \citep{Hutsemekers:2000fv,Hutsemekers:2005iz}. In addition, the spatial variation of fine-structure constant \citep{Webb:2010hc,King:2012id,Mariano:2012wx,Molaro:2013saa} and MOND acceleration scale \citep{Zhou:2017lwy,Chang:2018vxs,Chang:2018lab} are inconsistent with isotropic universe. These phenomena may imply a preferred direction of the Universe and hence a violation of the cosmological principle.

As standard candles, the supernovae of type Ia (SNe Ia) have been used to investigate the accelerating expansion of the Universe  \citep{Riess:1998cb,Perlmutter:1998np}, and the possible deviations from the isotropic universe \citep{Schwarz:2007wf,Gupta2008,Antoniou:2010gw,Blomqvist:2010ky,Colin:2010ds,Cai:2011xs,Mariano:2012wx,Cai:2013lja,Kalus:2012zu,Zhao:2013yaa,Wang:2014vqa,Yang:2013gea,Chang:2014jza,Chang:2014wpa,Heneka:2013hka,Chang:2014nca,Bengaly:2015dza,Li:2015uda,Javanmardi:2015sfa,Lin:2015rza,Lin:2016jqp,Salehi:2015ira,Salehi:2016sta,Li:2017rqj,Ghodsi:2016dwp,Wang:2017ezt,Andrade:2017iam,Chang:2017bbi,Deng:2018yhb},
with the datasets given by the Union2 sample \citep{Amanullah:2010vv}, the Union2.1 sample \citep{Suzuki:2011hu} and the ``Joint Light-curve Analysis'' (JLA) sample \citep{Betoule:2014frx}. \citet{Antoniou:2010gw} first used the hemisphere comparison (HC) method to the Union2 sample and they found a certain cosmological preferred direction with maximum accelerating expansion rate. A similar preferred direction has been found in $\omega$CDM and CPL parameterized model by \citet{Cai:2011xs}. \citet{Mariano:2012wx} first used the dipole fitting (DF) method to the Union2 sample and they found the existence of dark energy dipole at $2\sigma$ level. \citet{Zhao:2013yaa} used the Hierarchical Equal Area isoLatitude Pixelation \footnote[1]{\url{https://healpix.sourceforge.io/}} \citep[HEALPix,][]{Gorski:2004by} to divide the Union2 SNe Ia into 12 subsets and a dipole of deceleration parameter has been found at more than $2\sigma$ level. Meanwhile, a model-independent approach \citep{Schwarz:2007wf,Kalus:2012zu}  was used to estimate the cosmic anisotropy at low-redshift range. In addition, these works have been extended to include gamma-ray bursts (GRBs) \citep{Wang:2014vqa,Chang:2014jza} and the fine-structure constant \citep{Mariano:2012wx,Li:2015uda,Li:2017rqj}, and the preferred direction still exists. A similar cosmic anisotropy has been found in the Union2.1 sample \citep{Yang:2013gea,Javanmardi:2015sfa,Bengaly:2015dza,Li:2015uda,Lin:2016jqp}. Different with the Union2 or Union2.1 sample, the JLA sample doesn't show any convincing evidence for the existence of the cosmic anisotropy \citep{Bengaly:2015dza,Lin:2015rza,Chang:2017bbi,Wang:2017ezt,Deng:2018yhb,Sun:2018epo}.
Constraining the anisotropic amplitude and direction in $\Lambda$CDM, $\omega$CDM and CPL models with the JLA sample gave a zero result \citep{Lin:2015rza}. No significant deviation from the isotropic universe was found through a redshift tomographic analysis on the JLA sample in the dipole-modulated $\Lambda$CDM model \citep{Chang:2017bbi}. Constrained by the JLA sample, an anisotropic universe model with Bianchi-I metric was found to be consistent with the isotropic universe \citep{Wang:2017ezt}. No significant dipole anisotropy \citep{Deng:2018yhb,Sun:2018epo} was found in JLA sample by using the CosmoMC \citep{Lewis:2002ah}.

Recently, \citet{Scolnic:2017caz} released the Pantheon sample which includes 1048 spectroscopically confirmed SNe Ia. The Pantheon sample comprises 279 SNe Ia discovered by the Pan-STARRS1 (PS1) Medium Deep Survey, and SNe Ia from various Low-$z$, SDSS, SNLS and HST surveys. The redshifts of these SNe Ia are in the range $0.01 < z < 2.26$. Compared to previous SNe Ia sample such as Union2 \citep{Amanullah:2010vv}, Union2.1 \citep{Suzuki:2011hu} and JLA \citep{Betoule:2014frx}, the number of SNe Ia in Pantheon sample is enlarged and all subsamples are cross-calibrated so that systematic uncertainties have been reduced. Therefore, the Pantheon sample is an ideal object to investigate the cosmic anisotropy. However, the cosmic anisotropy has not been found in the Pantheon sample according to some researches \citep{Andrade:2018eta,Deng:2018jrp,Sun:2018cha,Li:2019bqb}. \citet{Deng:2018jrp} employed the HC method, the DF method and HEALPix to the Pantheon sample, and they found null signal for the cosmic anisotropy. \citet{Sun:2018cha} also used  the HC method and DF method to the Pantheon sample, and they found the cosmic anisotropy is weakly dependent on redshift and the isotropic cosmological model is an excellent approximation. Meanwhile, \citet{Andrade:2018eta} used a model-independent analysis by selecting low-redshift subsamples, and they found the current SN Ia data favour the hypothesis of cosmic isotropy. In fact, the distribution of Pantheon SNe Ia is inhomogeneous in the sky, which could bring significant impact on the cosmic anisotropy. There are large distinctions between the distributions of each subsample, for example, the Low-$z$ subsample is the most homogeneous while the PS1, SDSS and SNLS subsamples are extremely inhomogeneous in the sky. The distribution-dependence of methods that test the cosmic anisotropy has been noticed \citep{Appleby:2013ida,Chang:2014nca,Jimenez:2014jma,Bengaly:2015dza,Lin:2016jqp,Andrade:2018eta,Sun:2018cha}. \citet{Chang:2014nca} and \citet{Lin:2016jqp} suggested that the HC method strongly depends on the distribution of SNe Ia in the sky. \citet{Jimenez:2014jma} found the preferred direction in the Union2 is aligned with the orthogonal direction of the SDSS observational plane. \citet{Bengaly:2015dza} found that the dipolar direction of the Hubble map can be attributed to the intrinsic anisotropy of the JLA sample. It is worth investigating the impact of inhomogeneous distribution of Pantheon SNe Ia on the cosmic anisotropy.

As mentioned before, there are large distinctions between the distributions of each subsample. In order to investigate the impact of each subsample on the HC method and DF method, we will exclude individual subsamples from the Pantheon sample in turn. The full Pantheon sample and five combinations among Pantheon are used to test the cosmic anisotropy by performing the HC method and DF method, respectively. The rest of this paper is arranged as follows: In section \ref{sec:pantheon}, we introduce the Pantheon sample and its distribution in the sky. In section \ref{sec:HC}, we use the HC method to investigate the cosmic anisotropy. In section \ref{sec:DF}, we use the DF method to investigate the cosmic anisotropy. Finally, conclusions are given in section \ref{sec:conclusion}.

\begin{figure*}
	\centering
	\subfigure[Pantheon]{
		\includegraphics[width=8cm]{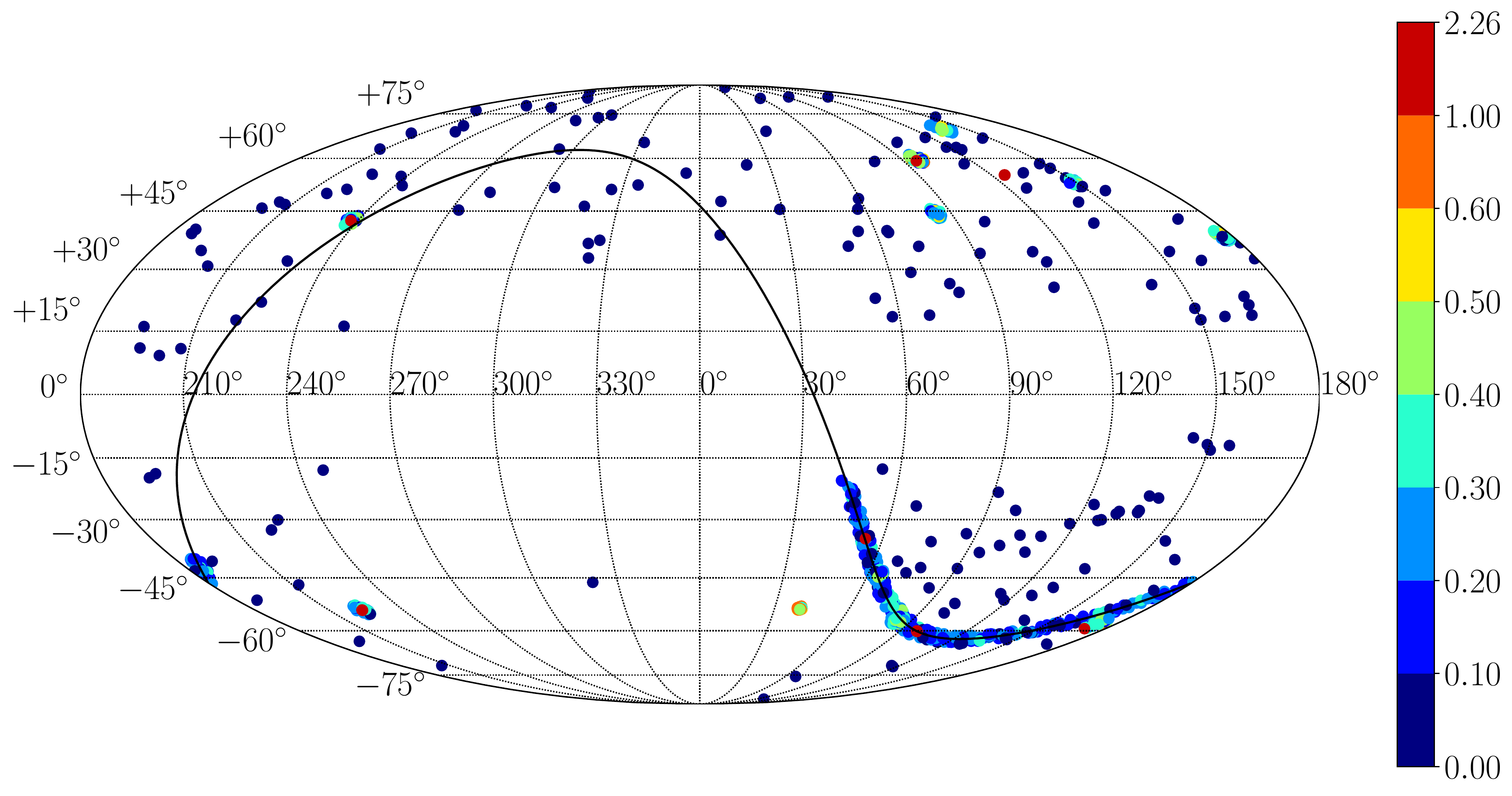} }
	\quad
	\subfigure[Low-$z$]{
		\includegraphics[width=8cm]{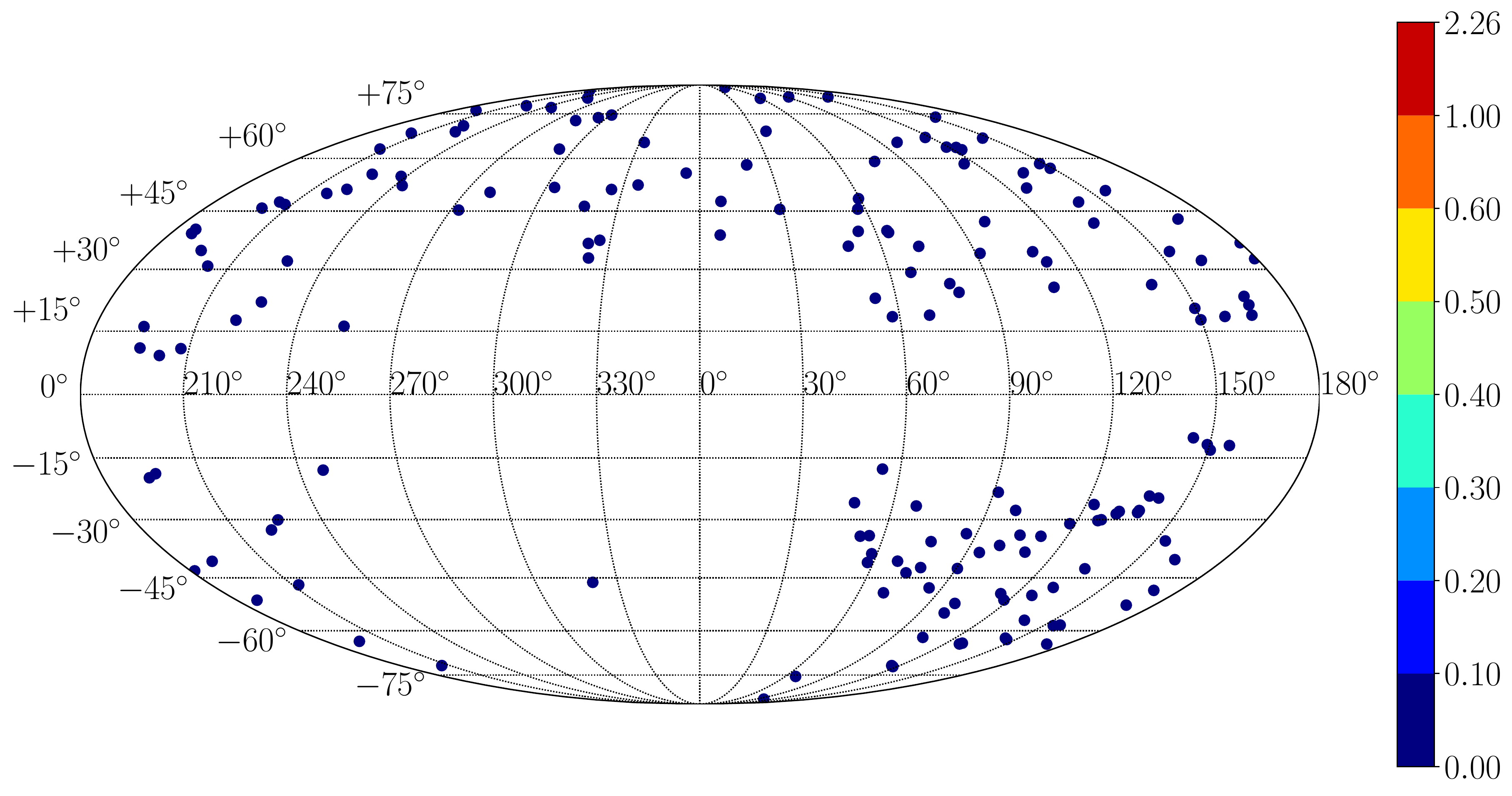} }
	\quad
	\subfigure[PS1]{
		\includegraphics[width=8cm]{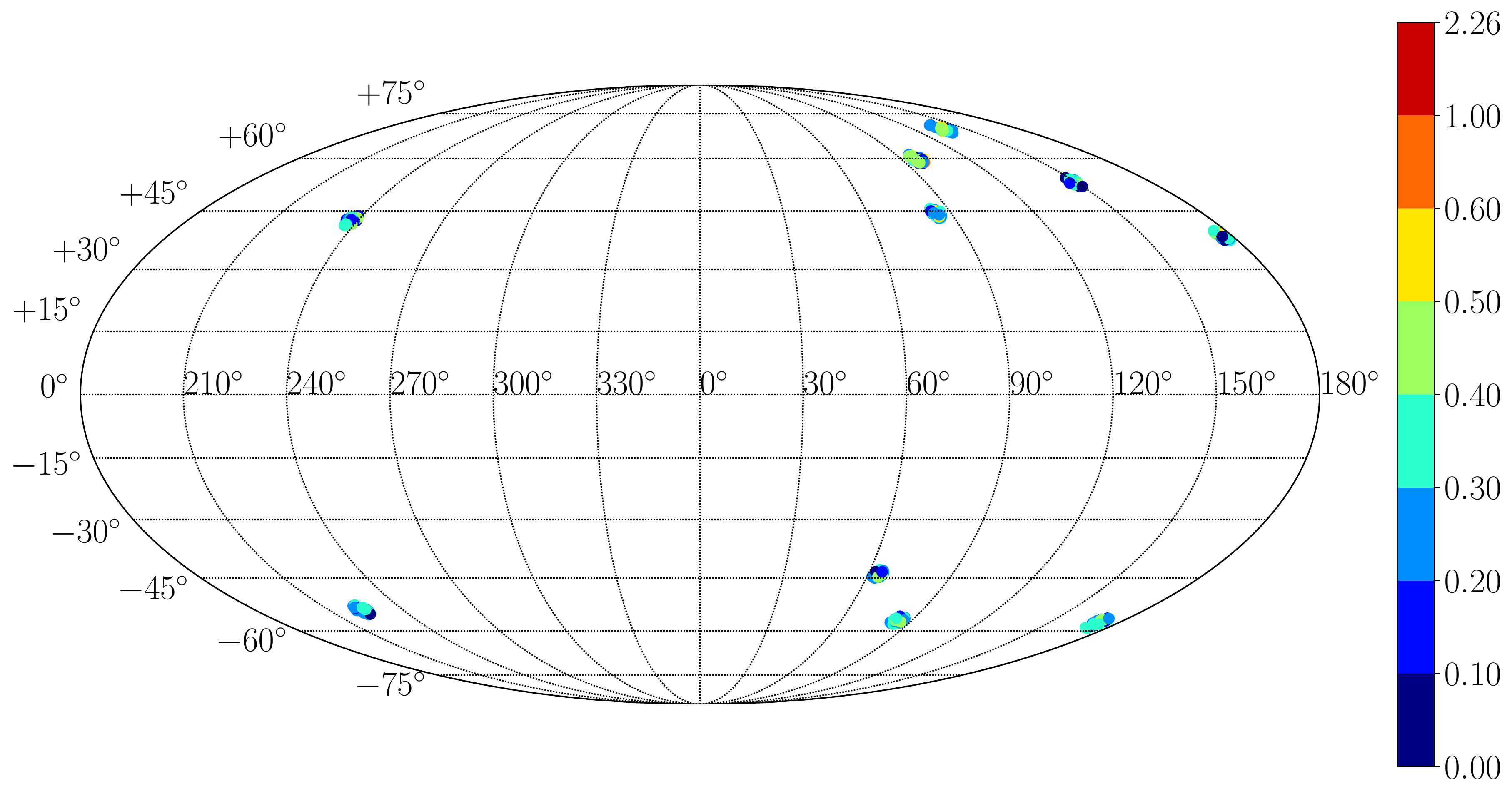} }
	\quad
	\subfigure[SDSS]{
		\includegraphics[width=8cm]{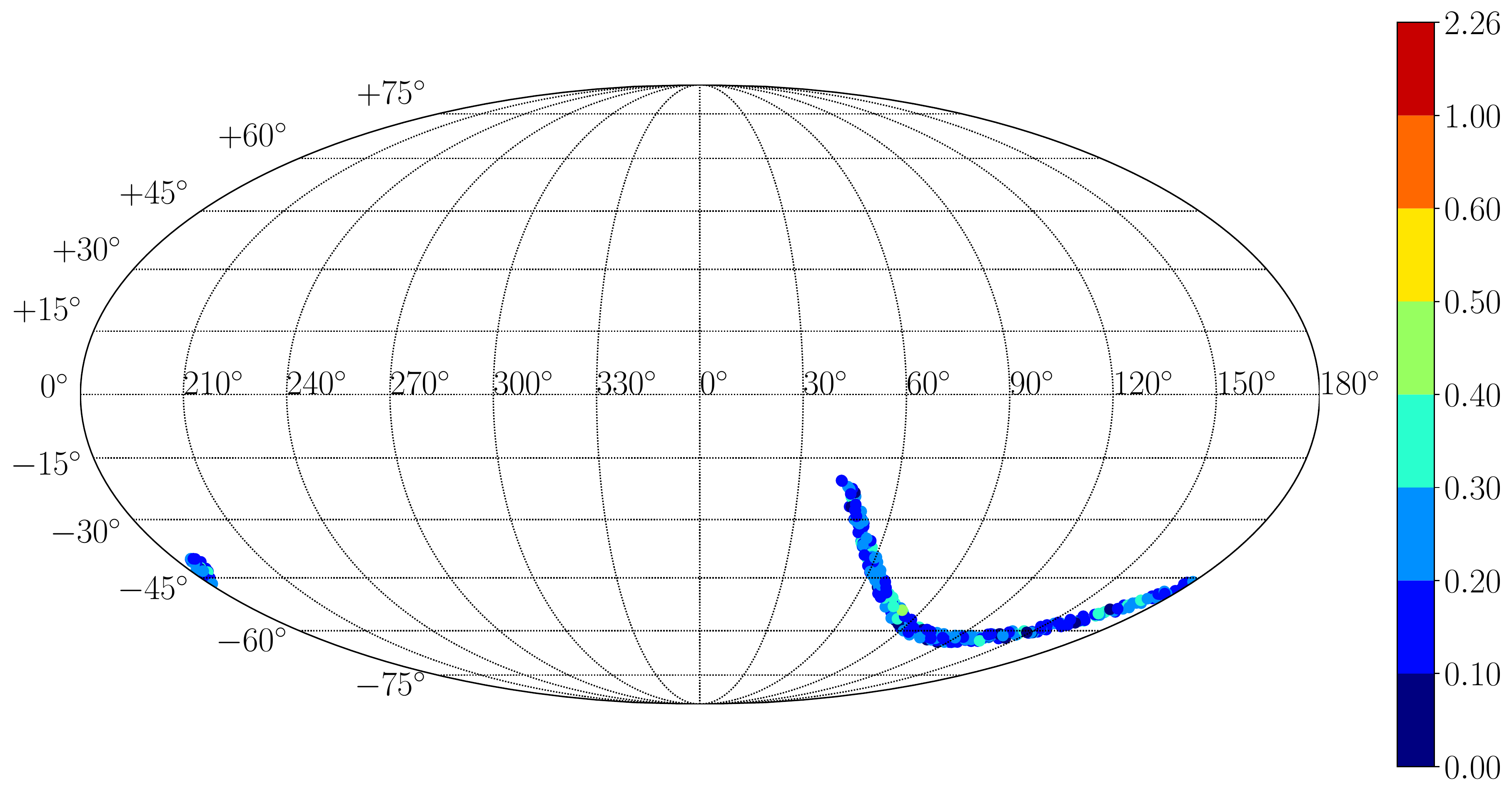} }
	\quad
	\subfigure[SNLS]{
		\includegraphics[width=8cm]{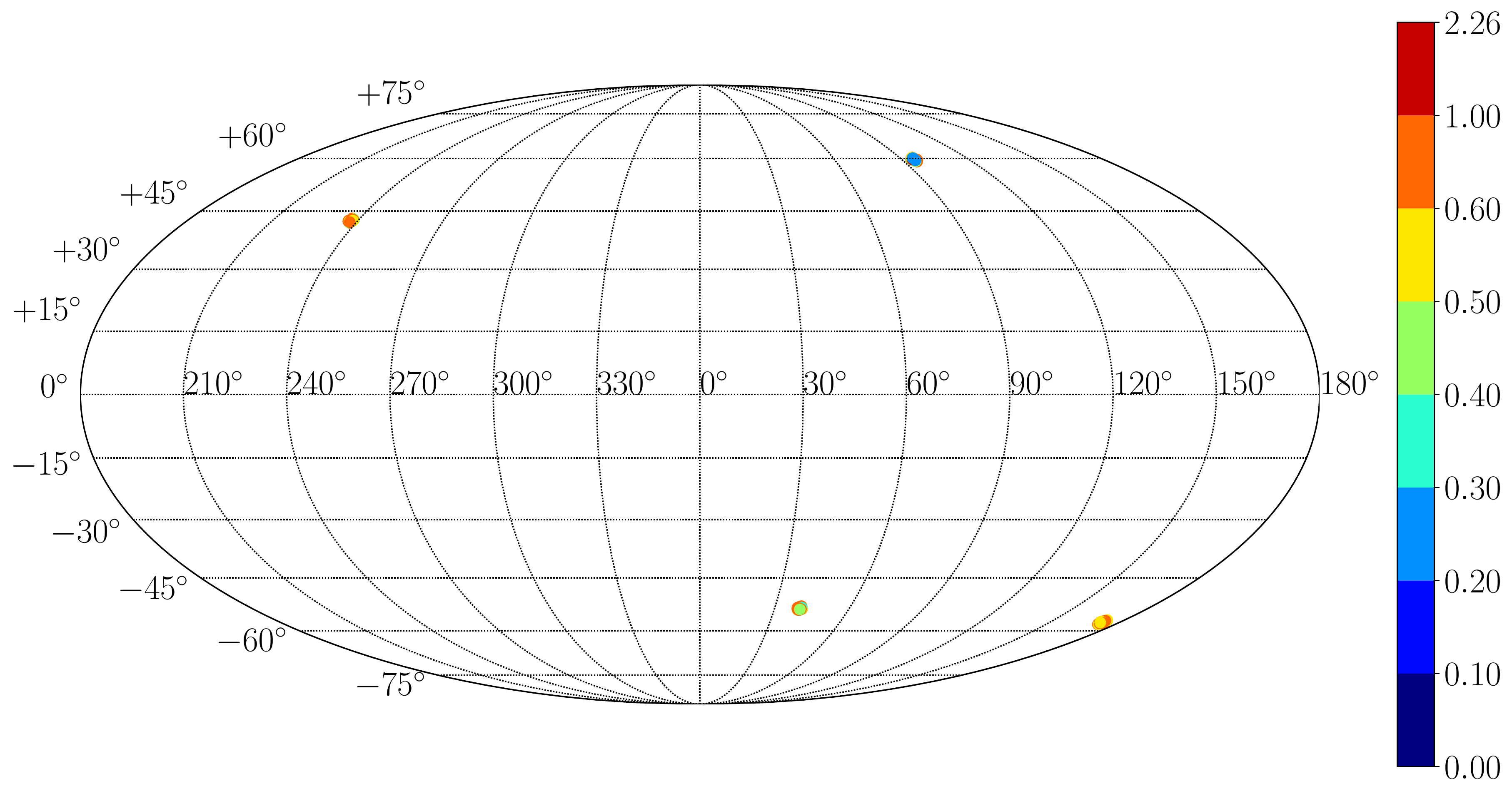} }
	\quad
	\subfigure[HST]{
		\includegraphics[width=8cm]{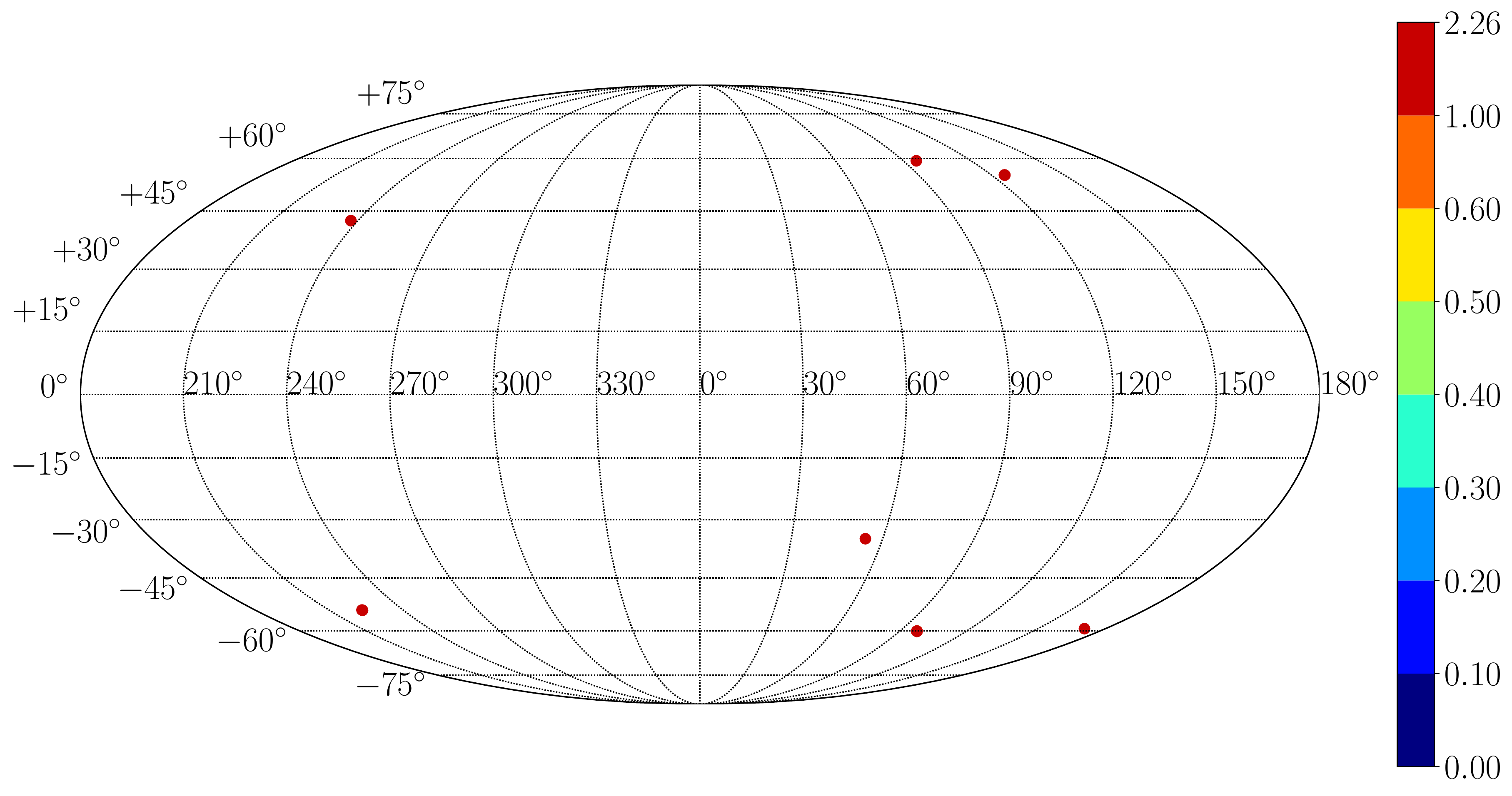} }
	\caption{ The distribution of 1048 Pantheon SNe Ia in the galactic coordinate system. The distribution of each subsample of Pantheon is also shown. The pseudo-colors represent the redshift of these SNe Ia. The black solid curve represents the celestial equator.}
	\label{fig:distribution}
\end{figure*}

\section{Pantheon Sample}
\label{sec:pantheon}
\begin{figure}
	\begin{center}
		\includegraphics[width=8cm]{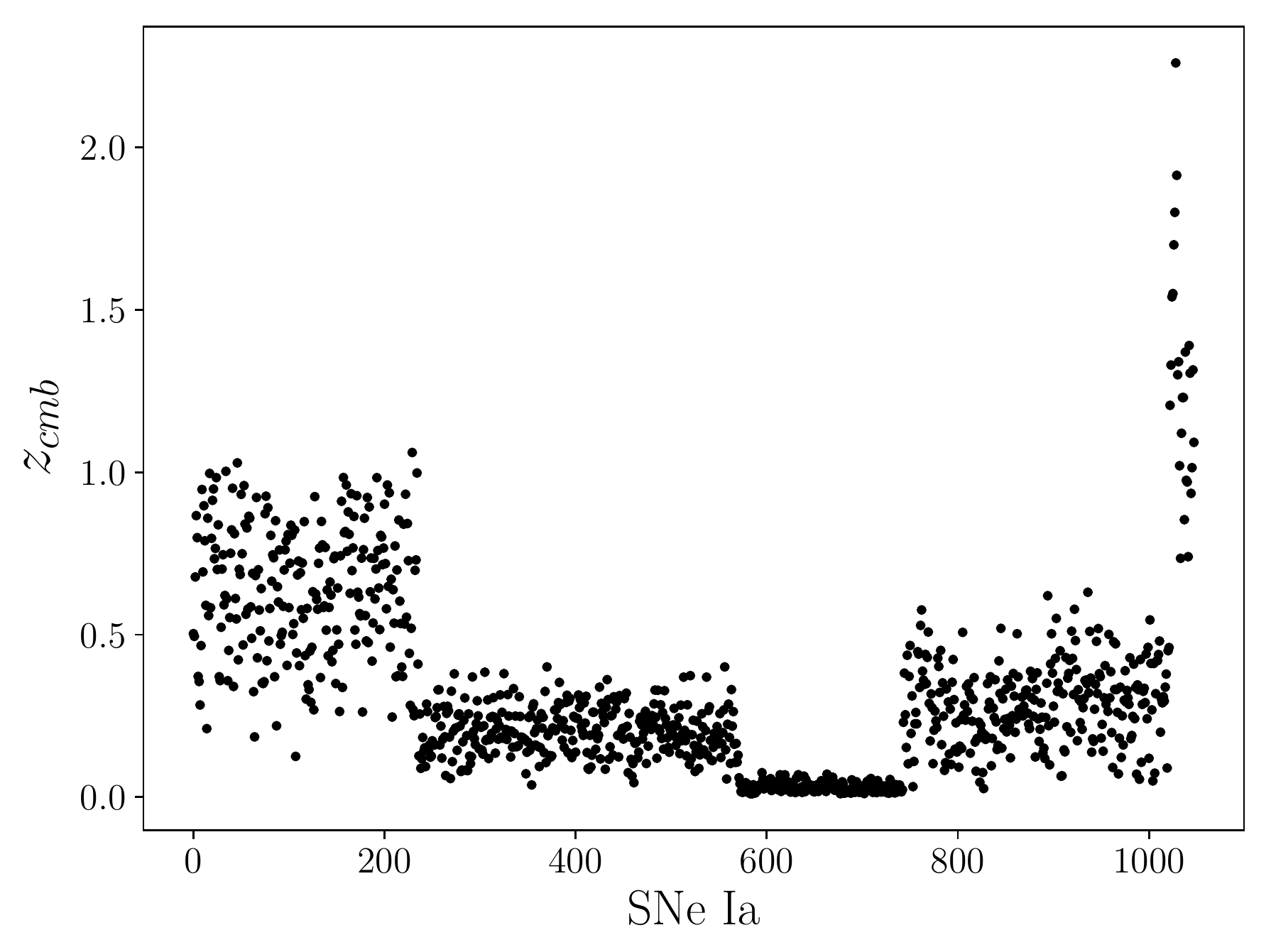}
		\caption{The redshift distributions of 1048 Pantheon SNe Ia. Each subsample of Pantheon has different redshift range. From left to right, the subsamples are SNLS, SDSS, Low-$z$, PS1 and HST.}
		\label{fig:redshift}
	\end{center}
\end{figure}

\noindent
\citet{Scolnic:2017caz} compiled the Pantheon sample which includes 1048 spectroscopically confirmed SNe Ia. The Pantheon sample comprises 279 SNe Ia discovered by the Pan-STARRS1 (PS1) Medium Deep Survey, and SNe Ia from Low-$z$, SDSS, SNLS and HST subsamples, where Low-$z$ is the compilation of all the smaller low-$z$ surveys and HST is the compilation of all the HST surveys. The redshifts of these SNe Ia are in the range $0.01<z<2.26$. In Fig. \ref{fig:distribution}, we display the distribution of these SNe Ia in the sky of the galactic coordinate system. As can be seen, the Pantheon SNe Ia are not uniformly distributed in the sky, half of them are located in the galactic south-east. The pseudo-colors indicate the redshift of these SNe Ia, which is also shown in Fig. \ref{fig:redshift}. Each subsample of Pantheon has different redshift range. We also display the sky distribution for each subsample of Pantheon in Fig. \ref{fig:distribution}. It is found that three subsamples are extremely inhomogeneous in the sky. For the PS1 subsample, 279 SNe Ia cluster in ten directions, and each direction has nearly 30 SNe Ia. For the SDSS subsample, 335 SNe Ia cluster in a narrow strip which corresponds to the equator of the equatorial coordinate system. For the SNLS subsample, 236 SNe Ia cluster in four directions, and each direction has nearly 60 SNe Ia. Due to the cosmic anisotropy is sensitive to the spatial distribution, therefore these inhomogeneous distributions of SNe Ia could bring significant impact on the cosmic anisotropy. In this paper, we will employ the HC method and DF method to investigate the anisotropy of the Universe with the Pantheon sample. In order to investigate the impact of each subsample among Pantheon on the overall anisotropy, we exclude individual subsamples from the full Pantheon sample in turn. This method insures enough SNe Ia used in our analysis. The full Pantheon sample and its five combinations are listed as follows,
\begin{enumerate}[(1)]
	\item Full Pantheon
	\item Pantheon without Low-$z$
	\item Pantheon without PS1
	\item Pantheon without SDSS
	\item Pantheon without SNLS
	\item Pantheon without HST
\end{enumerate}

\citet{Scolnic:2017caz} employed the SALT2 \citep{Guy:2010bc} light-curve fitter to determine the observed distance modulus by using a modified version of the Tripp formula \citep{Tripp:1998},
\begin{equation}
\mu_{obs}=m_B-M+\alpha x_1-\beta c+\Delta_M+\Delta_B,
\end{equation}
where $\mu_{obs}$ is the observed distance modulus, $m_B$ is the apparent $B$-band magnitude and $M$ is the absolute $B$-band magnitude of a fiducial SNe Ia with $x_1=0$ and $c=0$. $x_1$ is the stretch parameter and  $\alpha$ is the coefficient of the relation between luminosity and stretch. $c$ is the color parameter and $\beta$ is the coefficient of the relation between luminosity and color. $\Delta_M$ is a distance correction based on the host galaxy mass of the SN and $\Delta_B$ is a distance correction based on predicted biases from simulations. Since the absolute magnitude of an SN Ia is degenerated with the Hubble constant, only the corrected apparent magnitudes are given in \citet{Scolnic:2017caz}, i.e. $m_{obs}=\mu_{obs}+M$. Then one can use these data to constrain the cosmological parameters.

In the standard cosmological model, the theoretical distance modulus takes the form,
\begin{equation}
\mu_{th}=5\,\log_{10}\frac{\mathit{d_L}}{\mathrm{Mpc}}+25,
\end{equation}
where $d_L=(c/H_0)D_L$ is the luminosity distance, $H_0$ is the Hubble constant, $c$ is the speed of light,
\begin{equation}
\label{lumdis}
D_L=(1+z_{hel})\int_0^{z_{cmb}}\frac{dz}{E(z)},
\end{equation}
here $z_{hel}$ is the heliocentric redshift and $z_{cmb}$ is the CMB frame redshift. In the flat $\Lambda$CDM model, the dimensionless Hubble parameter is given as
\begin{equation}
E(z)=\sqrt{\Omega_{m}(1+z)^3+(1-\Omega_{m})},
\end{equation}
where $\Omega_{m}$ denotes the matter density at the present epoch. Then we can compare the observed apparent magnitude $m_{obs}$ with the theoretical apparent magnitude $m_{th}$, and the latter is given as
\begin{equation}\label{eq:mth}
m_{th}=\mu_{th}+M = 5\,\log_{10}D_{L}+\mathcal{M},
\end{equation}
where the nuisance parameter $\mathcal{M}$ depends on the absolute magnitudes $M$, and the Hubble constant $H_0$.

To infer the best-fitting values of $\Omega_{m}$ and $\mathcal{M}$, we employ the $\chi^2$ statistic,
\begin{equation}
\chi^2=\Delta \bm{\mu}^T \cdot \bm{\mathit{C}}^{-1} \cdot \Delta \bm{\mu}=\Delta \bm{\mathit{m}}^T \cdot \bm{\mathit{C}}^{-1} \cdot \Delta \bm{\mathit{m}},
\end{equation}
where $\Delta \bm{\mu}=\bm{\mu}_{obs}-\bm{\mu}_{th}$ or $\Delta \bm{\mathit{m}}=\bm{\mathit{m}}_{obs}-\bm{\mathit{m}}_{th}$. The total covariance matrix $\bm{\mathit{C}}$ is given as
\begin{equation}
\bm{\mathit{C}}=\bm{\mathit{D}}_{stat}+\bm{\mathit{C}}_{sys},
\end{equation}
where the diagonal matrix $\bm{\mathit{D}}_{stat}$ represents the statistical uncertainties and the covariance matrix $\bm{\mathit{C}}_{sys}$ denotes systematic uncertainties. The best-fitting parameters can be obtained by minimizing $\chi^2$.

\section{Hemisphere comparison method and result}\label{sec:HC}\noindent
The HC method has been widely used to investigate the cosmic anisotropy. For example, it has been used for investigating the anisotropic accelerating expansion of Universe \citep{Schwarz:2007wf,Antoniou:2010gw},  the hemispherical power asymmetry of CMB temperature anisotropies \citep{Ade:2013nlj,Akrami:2014eta} and a varying MOND acceleration scale \citep{Zhou:2017lwy}. We can look for a preferred axis to describe the cosmic anisotropy by the HC method. The main steps of the HC method are as follows:
\begin{enumerate}[(i)]
	\item Generate an arbitrary direction $\bm{\hat{n}}(l,b)$ in the galactic coordinate system, where $l$ and $b$ are longitude and latitude, respectively. Divide the celestial sphere into ``up'' and ``down'' hemispheres according the direction $\bm{\hat{n}}$.
	\item Split the Pantheon dataset into two subsets according to the position of SNe Ia in the sky. One subset is located in ``up'' hemisphere and another subset is located in ``down'' hemisphere.
	\item Find the best-fitting value of $\Omega_m$ and its $1\sigma$ error on each hemisphere. Define the anisotropy level (AL) by the normalized difference as
	\begin{equation}\label{eq:AL}
	\mathrm{AL} \equiv \frac{\Delta\Omega_{m}}{\bar{\Omega}_{m}}=2\cdot \frac{\Omega_{m,u}-\Omega_{m,d}}{\Omega_{m,u}+\Omega_{m,d}}.
	\end{equation}
	The $1\sigma$ error of AL is given by
	\begin{equation}\label{eq:ALsigma}
	\mathrm{\sigma_{AL}} = \frac{\sqrt{\sigma^2_{\Omega_{m,u}}+\sigma^2_{\Omega_{m,d}}}}{\Omega_{m,u}+\Omega_{m,d}}.
	\end{equation}
	\item Repeat for adequate directions, find the maximum anisotropy level and its corresponding direction.
\end{enumerate}

\begin{table*}
	\caption{The results of the HC method for the Pantheon sample and five combinations among Pantheon. We list the maximum AL and its corresponding direction $(l,b)$ in the galactic coordinate system, and the best-fitting values of $\Omega_m$ in ``up'' and ``down'' hemispheres, as well as the total number $N$ of SNe Ia and the number in each hemisphere.}
	\setlength{\tabcolsep}{2.4mm}{
		\begin{tabular}{lcccccccc}
			\hline
			\multicolumn{1}{l}{Sample}                   &$N$   &$N_u$ &$N_d$ &$\Omega_{m,u}$   &$\Omega_{m,d}$   & $\mathrm{AL}_{max}$ &$l[^{\circ}]$                &$b[^{\circ}]$               \\
			\hline
			\multicolumn{1}{l}{Full Pantheon}            &1048  &883   &165   &$0.313\pm0.023$  &$0.217\pm0.029$  &$0.361\pm0.070$      &$123.05^{+11.25}_{  -4.22}$  &$  4.78^{ +1.80}_{ -8.36}$  \\
			\multicolumn{1}{l}{Pantheon without Low-$z$} &876   &456   &420   &$0.361\pm0.033$  &$0.279\pm0.027$  &$0.257\pm0.067$      &$ 98.44^{+37.27}_{ -99.14}$  &$ 26.61^{+17.59}_{-30.79}$  \\
			\multicolumn{1}{l}{Pantheon without PS1}     &769   &669   &100   &$0.305\pm0.024$  &$0.199\pm0.030$  &$0.420\pm0.076$      &$123.05^{+10.55}_{  -4.92}$  &$  4.78^{ +1.80}_{ -7.77}$  \\
			\multicolumn{1}{l}{Pantheon without SDSS}    &713   &548   &165   &$0.304\pm0.025$  &$0.217\pm0.029$  &$0.333\pm0.073$      &$123.05^{+23.20}_{ -36.56}$  &$  4.78^{+15.96}_{-28.75}$  \\
			\multicolumn{1}{l}{Pantheon without SNLS}    &812   &627   &185   &$0.347\pm0.031$  &$0.215\pm0.041$  &$0.472\pm0.092$      &$103.36^{+37.97}_{ -35.16}$  &$-28.63^{+35.21}_{ -0.68}$  \\
			\multicolumn{1}{l}{Pantheon without HST}     &1022  &861   &161   &$0.305\pm0.024$  &$0.214\pm0.031$  &$0.350\pm0.076$      &$123.75^{+10.55}_{  -2.81}$  &$  4.18^{ +2.40}_{ -7.76}$  \\
			\hline
	\end{tabular}}
	\label{table:HC}
\end{table*}

\begin{figure*}
\centering
\subfigure[Full Pantheon]{
\includegraphics[width=8cm]{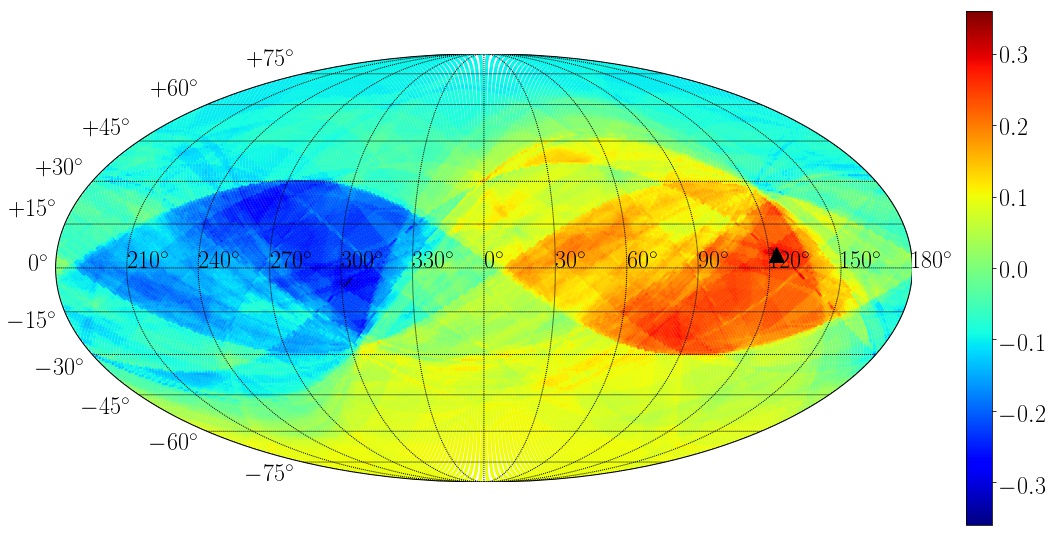} }
\quad
\subfigure[Pantheon without Low-$z$]{
\includegraphics[width=8cm]{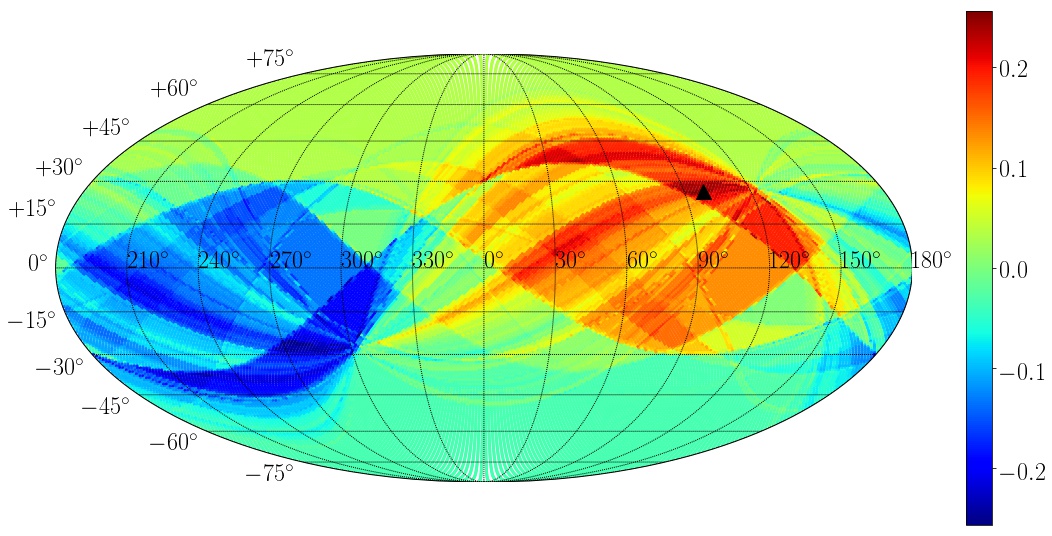} }
\quad
\subfigure[Pantheon without PS1]{
\includegraphics[width=8cm]{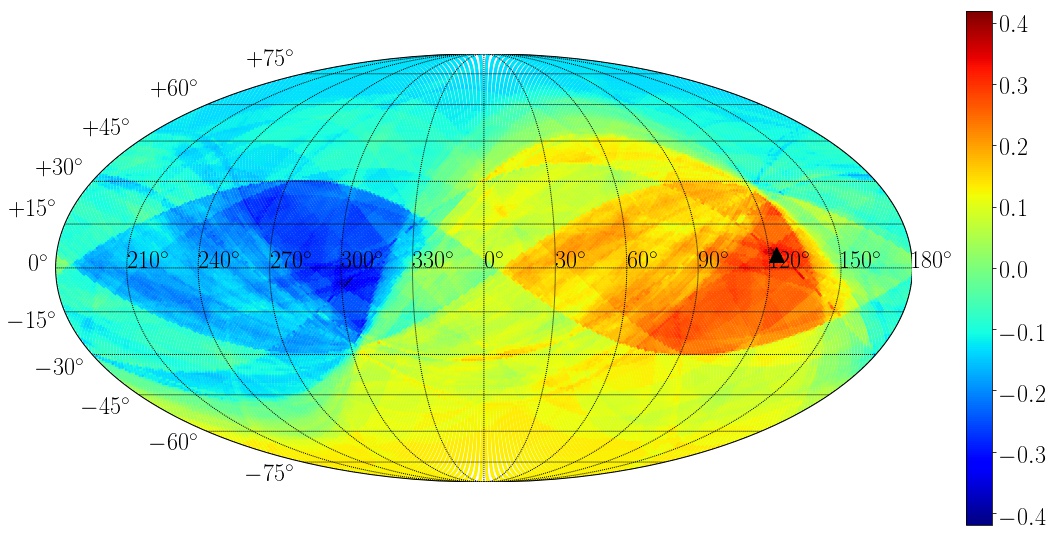} }
\quad
\subfigure[Pantheon without SDSS]{
\includegraphics[width=8cm]{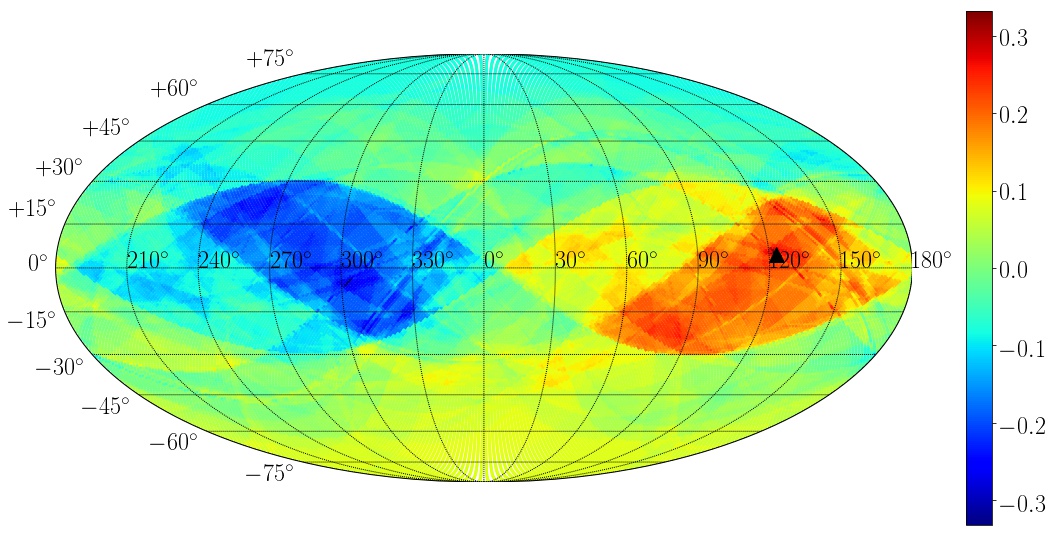} }
\quad
\subfigure[Pantheon without SNLS]{
\includegraphics[width=8cm]{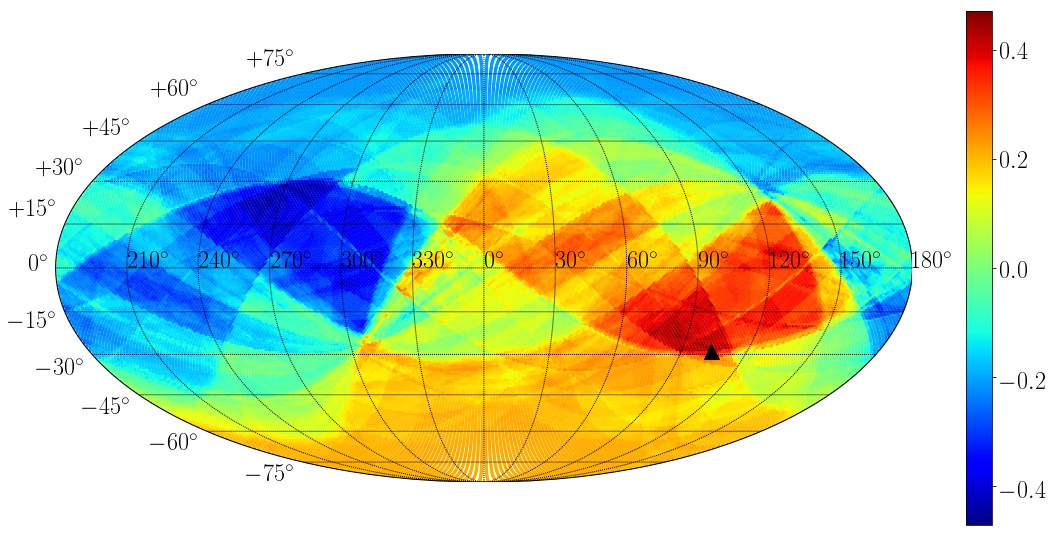} }
\quad
\subfigure[Pantheon without HST]{
\includegraphics[width=8cm]{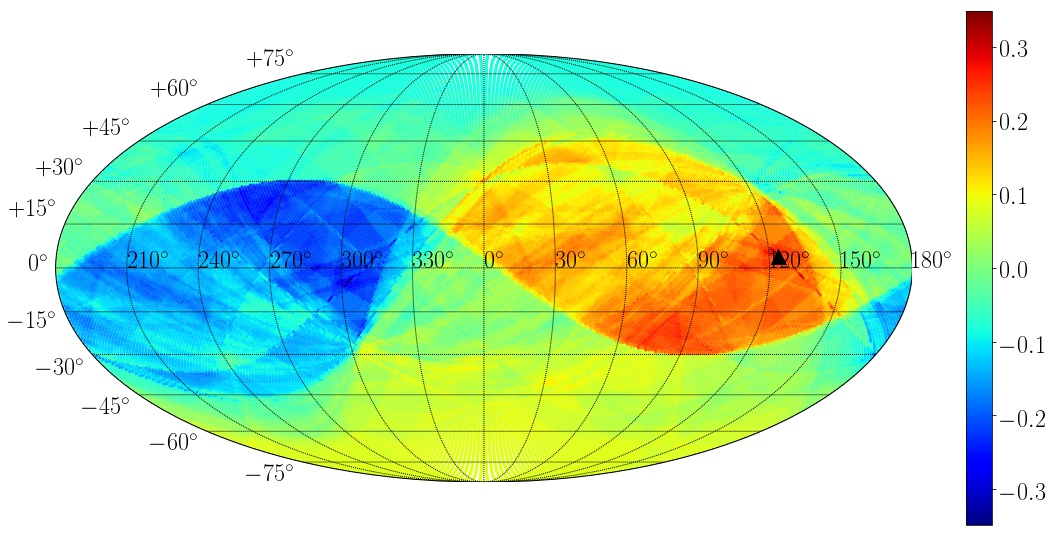} }
\caption{The pseudo-color map of the anisotropy level in the galactic coordinate system, derived from the HC method to the full Pantheon sample and five combinations among Pantheon. The triangle marks the maximum anisotropy level in the sky.}
\label{fig:HC}
\end{figure*}

\begin{figure*}
	\centering
	\subfigure[Full Pantheon]{
		\includegraphics[width=8cm]{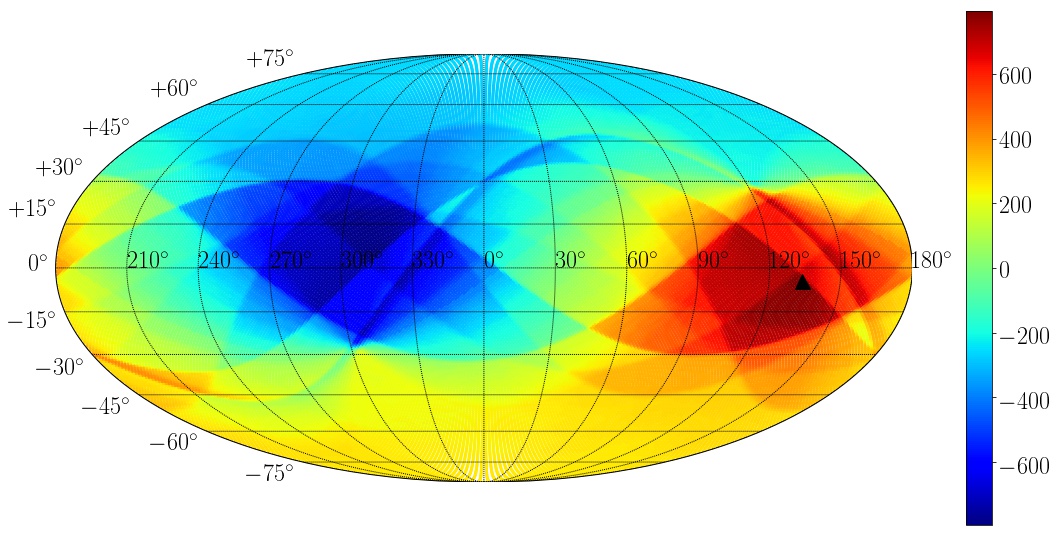} }
	\quad
	\subfigure[Pantheon without Low-$z$]{
		\includegraphics[width=8cm]{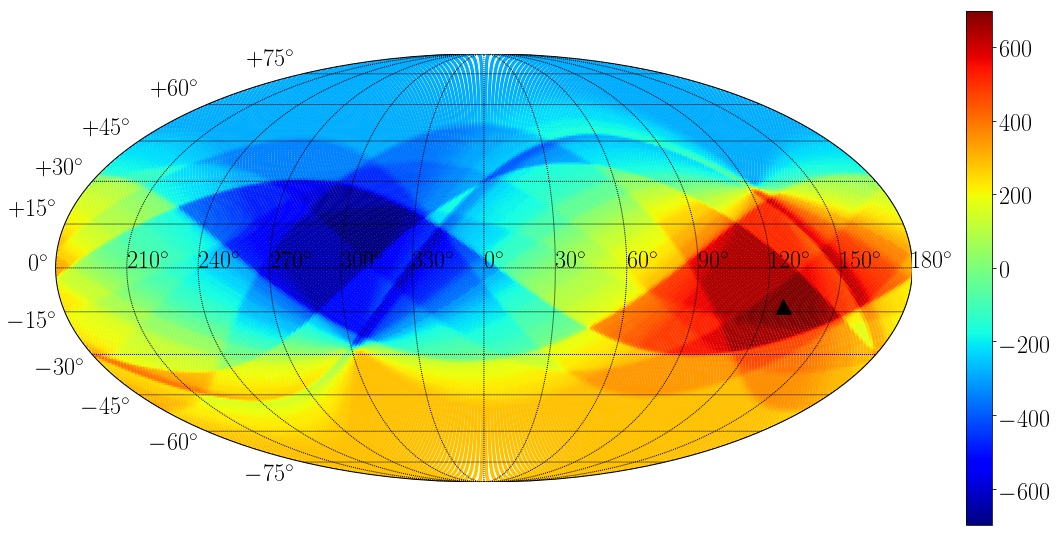} }
	\quad
	\subfigure[Pantheon without PS1]{
		\includegraphics[width=8cm]{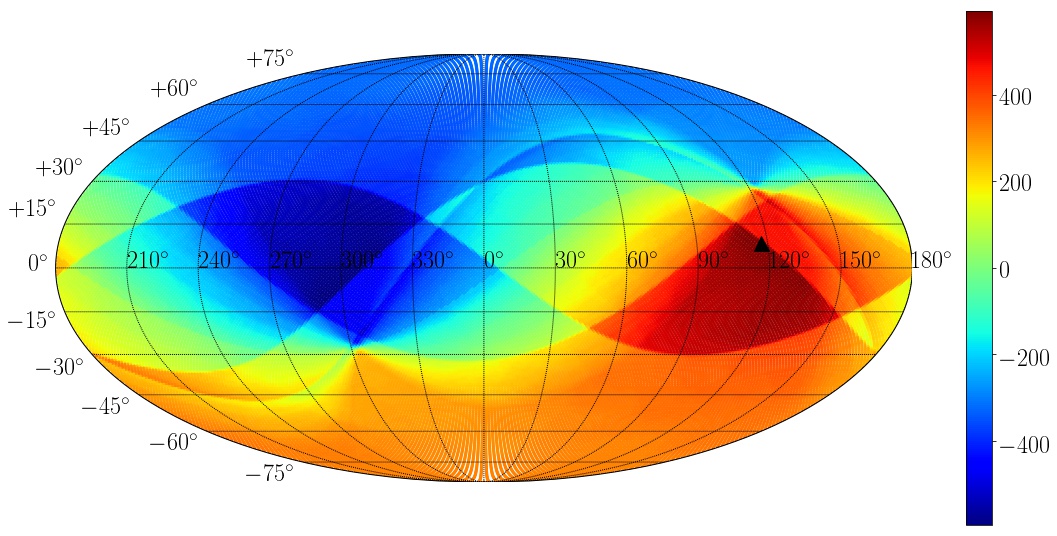} }
	\quad
	\subfigure[Pantheon without SDSS]{
		\includegraphics[width=8cm]{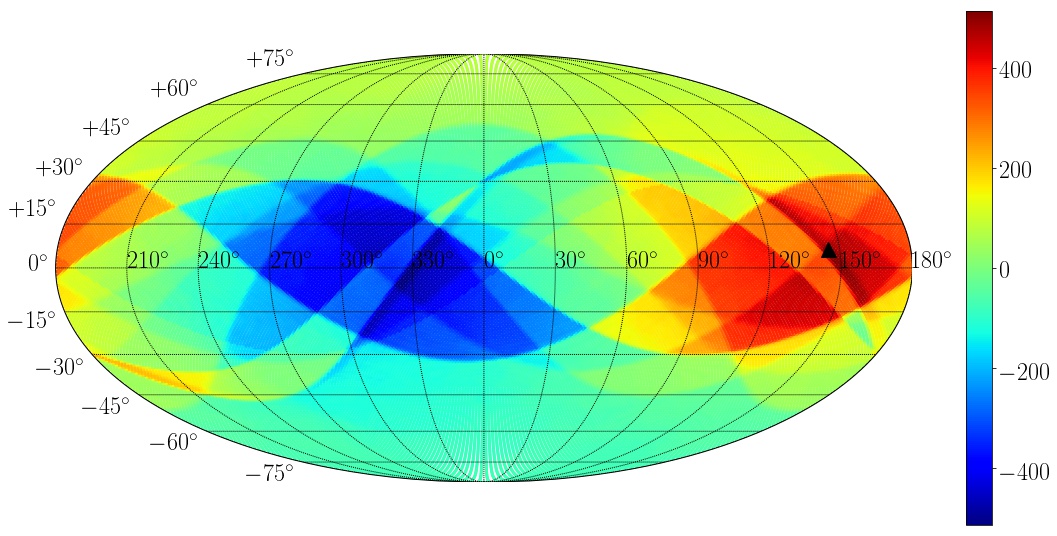} }
	\quad
	\subfigure[Pantheon without SNLS]{
		\includegraphics[width=8cm]{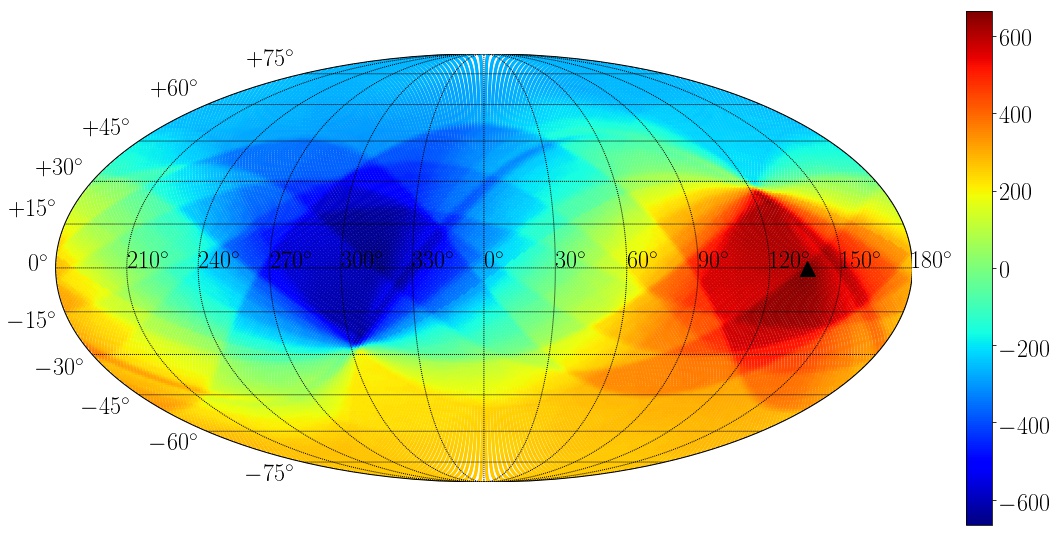} }
	\quad
	\subfigure[Pantheon without HST]{
		\includegraphics[width=8cm]{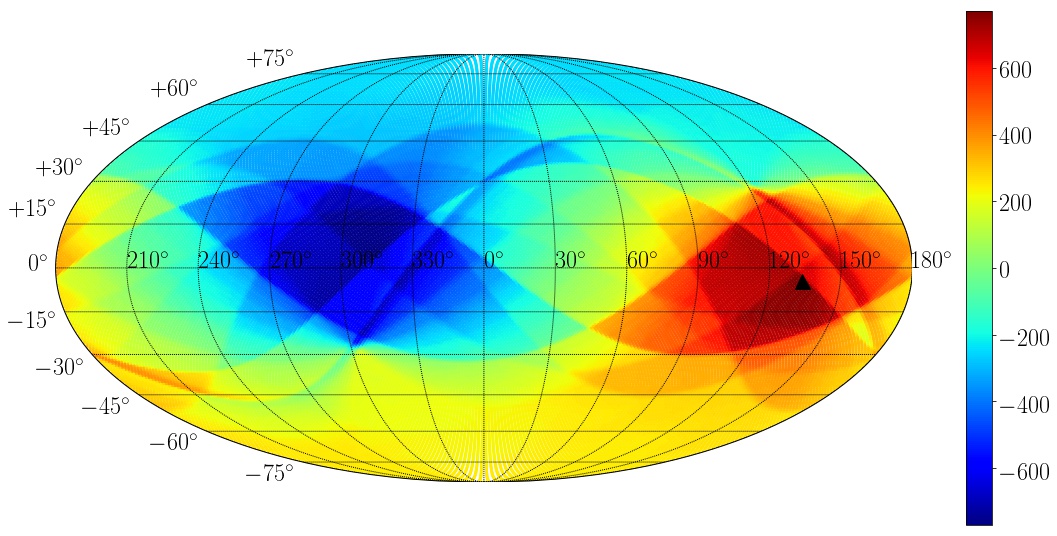} }
	\caption{The pseudo-color map of $\mathrm{AL}_N$ in the galactic coordinate system for the full Pantheon sample and five combinations among Pantheon. The triangle marks the maximum number difference in the sky.}
	\label{fig:Number}
\end{figure*}

The HEALPix package \citep{Gorski:2004by} is used to generate the directions $(l,b)$ in the HC method with $N_{side}=64$. The total number of the directions are $12\times N_{side}^2=49152$, which is more denser than that in \citet{Deng:2018jrp} and \citet{Sun:2018cha}. The high resolution could benefit for us to find the maximum in the distribution of AL.

At first, we employ the HC method to investigate the cosmic anisotropy by using the full Pantheon sample. The anisotropy level map is shown in Fig. \ref{fig:HC}. The triangle marks the maximum anisotropy level in the sky. Note that the anisotropy level map in Fig. \ref{fig:HC} is similar to  Fig. 2 in \citet{Deng:2018jrp} and Fig. 2 in \citet{Sun:2018cha}, but high resolution used in this paper benefits for us to find the maximum anisotropy level. We find the maximum anisotropy level is $\mathrm{AL}_{max}=0.361\pm0.070$ and the corresponding direction is $(l,b)=({123.05^{\circ}}^{+11.25^{\circ}}_{-4.22^{\circ}}, {4.78^{\circ}}^{+1.80^{\circ}}_{-8.36^{\circ}})$. The $1\sigma$ range of the maximum anisotropy direction is identified by the direction that corresponds to an anisotropy level within 1$\sigma$ from the maximum anisotropy level, i.e. $\mathrm{AL}=\mathrm{AL}_{max}\pm\sigma_{\mathrm{AL}}$. In the preferred direction, the best-fitting values of $\Omega_m$ in ``up'' and ``down'' hemispheres are $\Omega_{m,u=}0.313\pm0.023$  and $\Omega_{m,d}=0.217\pm0.029$, respectively. Since the deceleration parameter $q_0=-1+3\,\Omega_m/2$ in the flat $\Lambda$CDM model, the minimum accelerating expansion of Universe is in the direction of maximum AL, while the maximum accelerating expansion is in the opposite direction. However, we find the numbers of SNe Ia in ``up'' and ``down'' hemispheres are 883 and 165, respectively. The hemispheric asymmetry of Pantheon SNe Ia is shown in Fig. \ref{fig:Number}. The inhomogeneous distribution of Pantheon SNe Ia in the sky could have significant impact on the anisotropy level in HC method. We will discuss this issue later. All related results are summarized in Table \ref{table:HC}. In addition, we also perform the HC method for the nuisance parameter $\mathcal{M}$. We find its maximum anisotropy level is less than $0.3\%$, showing that no obvious signal of anisotropy was found for $\mathcal{M}$ with the Pantheon sample.

\begin{figure}
	\begin{center}
		\includegraphics[width=8cm]{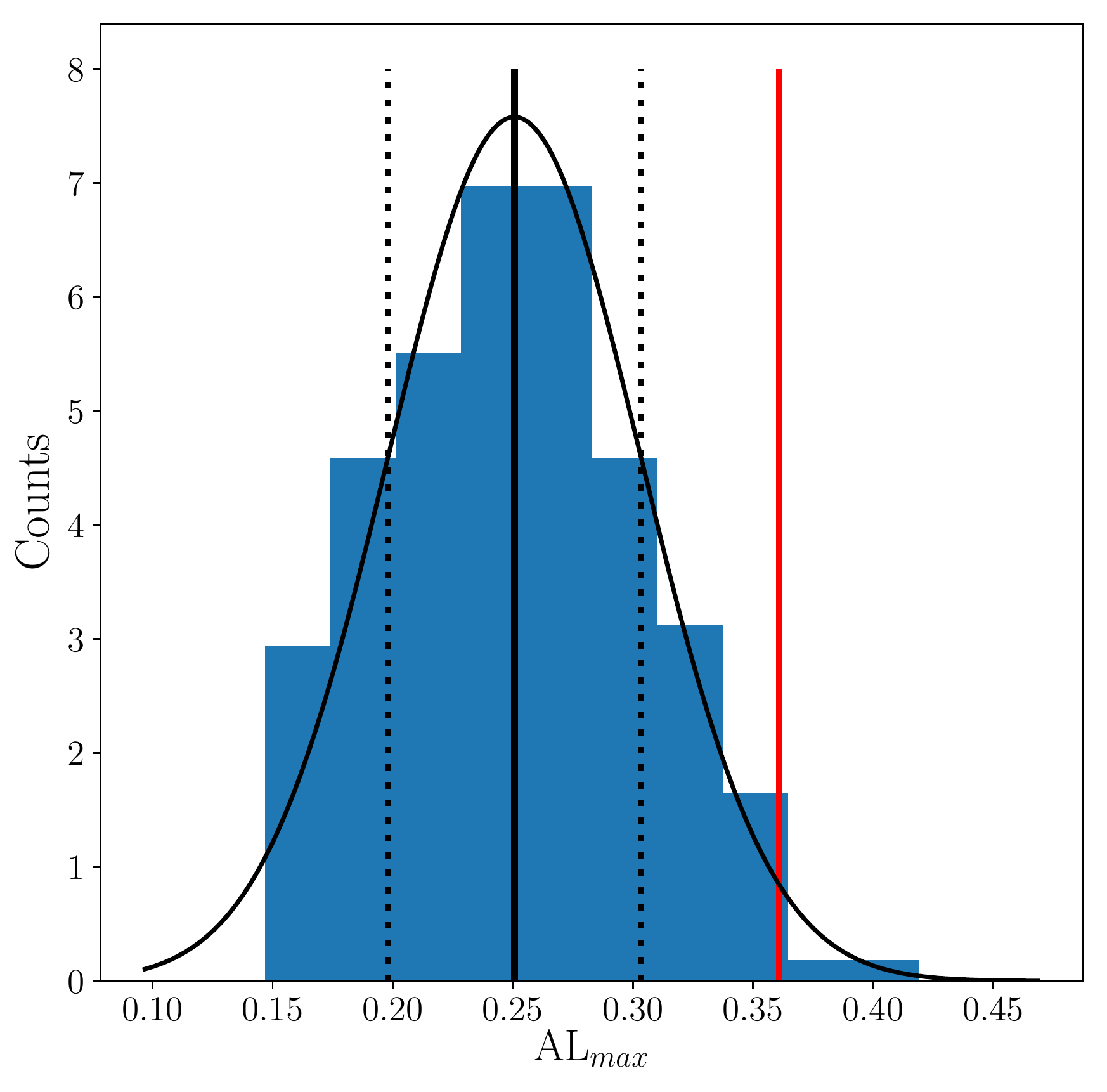}
		\caption{The distribution of the maximum anisotropy level in 200 simulated isotropic data sets, with black curve the best-fitting result to Gaussian function. Note that the counts have been normalized. The black solid and dashed vertical lines show the mean and the standard deviation, respectively. The red vertical line shows the maximum anisotropy level from the full Pantheon sample.}
		\label{fig:Simulation}
	\end{center}
\end{figure}

As a robust check, it is necessary to examine whether the maximum anisotropy level from the full Pantheon sample is consistent with statistical isotropy. The simulated isotropic data set is constructed by replacing the $i$th observed apparent magnitude $m_{obs,i}$ of the Pantheon sample by a random number from a Gaussian distribution with the mean determined by Eq. \eqref{eq:mth}, where $\Omega_m=0.298,~\mathcal{M}=23.809$ is the best fitting value to the full Pantheon sample, and the standard deviation is equal to the statistical uncertainty of $m_{obs,i}$. Then one can employ the HC method to this simulated isotropic data set and compare the maximum anisotropy level with that derived from the full Pantheon sample. Given the limitations of computing time, we construct 200 simulated isotropic data sets and they have acceptable statistics. The maximum anisotropy level from the HC method to these data sets is shown in Fig. \ref{fig:Simulation}. The distribution is fitted by a Gaussian function, the mean value is 0.251  and the standard deviation is 0.053. As can be seen, the maximum anisotropy level from the full Pantheon sample  is hardly reproduced by simulated isotropic data set and its statistical significance is about $2.1\sigma$.

\begin{figure}
	\begin{center}
		\includegraphics[width=8cm]{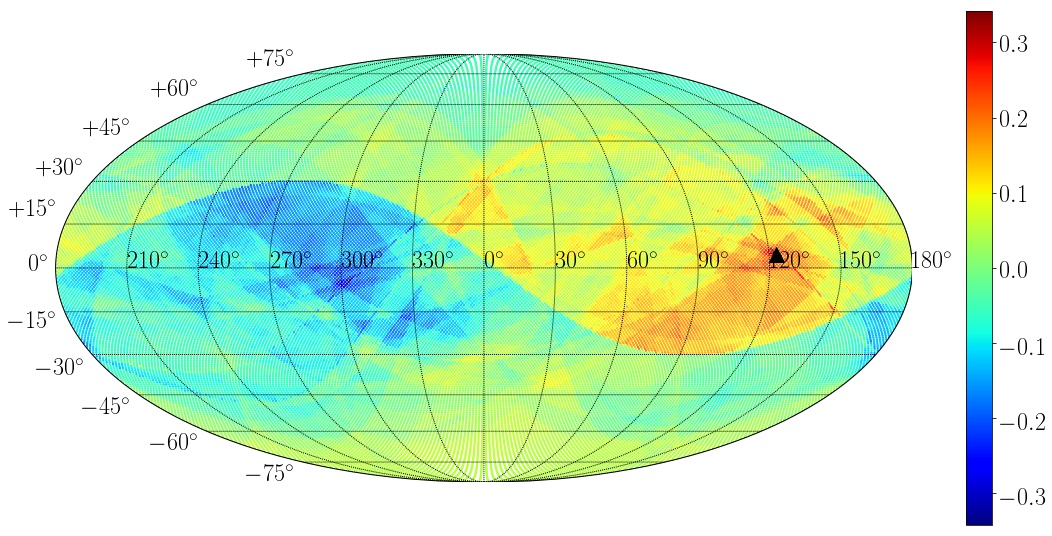}
		\caption{The pseudo-color map of the anisotropy level in the galactic coordinate system, derived from the HC method to the combination of Low-$z$ and  SNLS subsamples. The triangle marks the maximum anisotropy level in the sky.}
		\label{fig:HC2}
	\end{center}
\end{figure}

In order to investigate the impact of each subsample among Pantheon on the overall anisotropy in the full Pantheon, we exclude individual subsamples from the full Pantheon sample in turn and five combinations among Pantheon are used, which are listed in Sec. \ref{sec:pantheon}. We employ the HC method to these five combinations. The anisotropy level map for each combination is shown in Fig. \ref{fig:HC} and all related results are summarized in Table \ref{table:HC}. As can be seen, the AL map of the Pantheon without Low-$z$ or SNLS subsample has obvious difference with the AL map of the full Pantheon sample while other three combinations are consistent with the full Pantheon. It may imply that the Low-$z$ and SNLS subsamples have significant impact on the overall anisotropy in the full Pantheon. For the Pantheon without SNLS subsample, we find that the maximum anisotropy level is $\mathrm{AL}_{max}=0.472\pm0.092$ and the corresponding direction is $(l,b)=({103.36^{\circ}}^{+37.97^{\circ}}_{-35.16^{\circ}}, {-28.63^{\circ}}^{+35.21^{\circ}}_{-0.68^{\circ}})$. This direction is $38	.41^\circ$ away from the preferred direction in the full Pantheon. For the Pantheon without Low-$z$ subsample, we find that the maximum anisotropy level is $\mathrm{AL}_{max}=0.257\pm0.067$ and the corresponding direction is $(l,b)=({98.44^{\circ}}^{+37.27^{\circ}}_{-99.14^{\circ}}, {26.61^{\circ}}^{+17.59^{\circ}}_{-30.79^{\circ}})$. This direction is $32.07^\circ$ away from the preferred direction in the full Pantheon. Moreover, we employ the HC method to the combination of Low-$z$ and SNLS subsamples. We find a consistent preferred direction with that in the full Pantheon sample. The AL map is shown in Fig. \ref{fig:HC2}. The maximum anisotropy level is $\mathrm{AL}_{max}=0.342\pm0.090$ and the corresponding direction is $(l,b)=({123.05^{\circ}}^{+11.25^{\circ}}_{-4.92^{\circ}}, {4.78^{\circ}}^{+1.80^{\circ}}_{-8.36^{\circ}})$. These results confirm that the anisotropy in the  full Pantheon sample is mostly originating from the Low-$z$ and SNLS subsamples.

\begin{table}
	\caption{The maximum number difference and the number of SNe Ia in ``up'' and ``down'' hemisphere, and the corresponding direction for the full Pantheon sample and five combinations among Pantheon.}
	\setlength{\tabcolsep}{1.4mm}{
		\begin{tabular}{lcccrr}
			\hline
			\multicolumn{1}{l}{Sample}                   &$\mathrm{AL}_N$ &$N_u$ &$N_d$ &$l[^{\circ}]$  &$b[^{\circ}]$  \\
			\hline
			\multicolumn{1}{l}{Full Pantheon}            &796             &922   &126   &$133.95$       &$  -4.48$      \\
			\multicolumn{1}{l}{Pantheon without Low-$z$} &700             &788   &88    &$127.85$       &$ -13.01$     \\
			\multicolumn{1}{l}{Pantheon without PS1}     &595             &682   &87    &$117.07$       &$   8.39$      \\
			\multicolumn{1}{l}{Pantheon without SDSS}    &515             &614   &99    &$145.20$       &$   6.28$      \\
			\multicolumn{1}{l}{Pantheon without SNLS}    &666             &739   &73    &$135.70$       &$   0$         \\
			\multicolumn{1}{l}{Pantheon without HST}     &772             &897   &125   &$133.95$       &$  -4.48$      \\
			\hline
		\end{tabular}
		\label{table:Num}}
\end{table}

\begin{figure}
	\begin{center}
		\includegraphics[width=8cm]{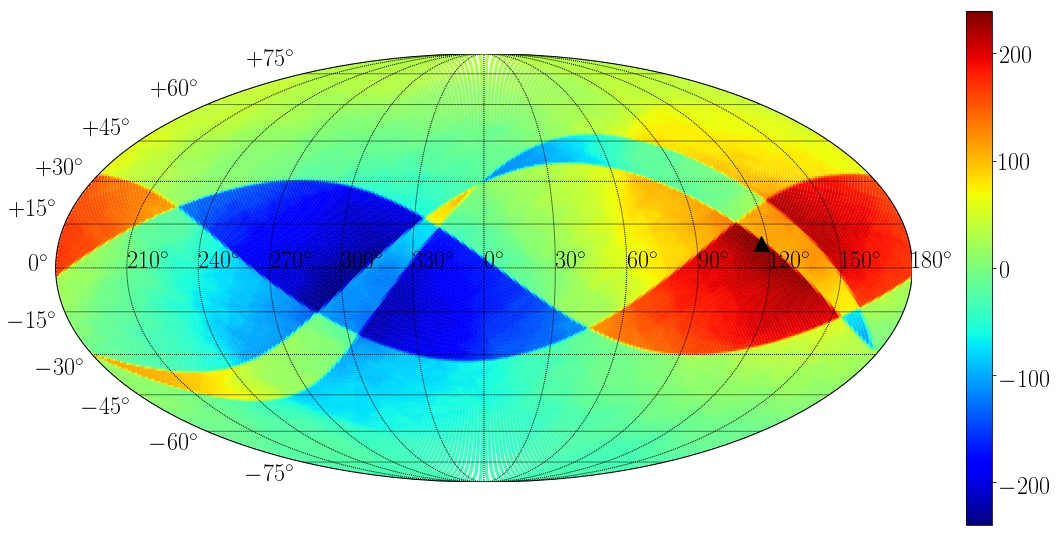}
		\caption{The pseudo-color map of $\mathrm{AL}_N$ in the galactic coordinate system for the combination of Low-$z$ and SNLS subsamples. The triangle marks the maximum number difference in the sky.}
		\label{fig:Number2}
	\end{center}
\end{figure}

For the full Pantheon sample and the combinations among Pantheon, we find the numbers of SNe Ia in ``up'' and ``down'' hemispheres have big difference in the maximum anisotropy direction. In order to investigate the
possible relation between the anisotropy level with the inhomogeneous distribution of SNe Ia in the galactic coordinate system, we define the number difference $\mathrm{AL}_N=N_u-N_d$ between the numbers of SNe Ia in ``up" and ``down" hemispheres, respectively. The distributions of $\mathrm{AL}_N$ for the full Pantheon sample and five combinations among Pantheon are shown in Fig. \ref{fig:Number}.  The maximum $\mathrm{AL}_N$ and the corresponding directions are listed in Table \ref{table:Num}. For the full Pantheon sample, the direction of the maximum $\mathrm{AL}_N$ is $(l,b)=({133.95^{\circ}}, {-4.48^{\circ}})$, and there are 922 SNe Ia in ``up'' hemisphere while only 126 SNe Ia in ``down'' hemisphere. In this direction, there are much more SNe Ia in ``up'' hemisphere than that in ``down'' hemisphere for each subsample. Even though individual subsamples have been excluded from the full Pantheon sample, there still be much more SNe Ia in the same ``up'' hemisphere due to the contributions of other four subsamples. This may be the reason why the direction of maximum $\mathrm{AL}_N$ in each combination is close to the direction in the full Pantheon. For the Pantheon without Low-$z$ (SNLS) subsample, the direction of maximum $\mathrm{AL}_N$ is only $10.44^\circ$ $(4.81^\circ)$ away from that in the full Pantheon. It is interesting to note that the maximum anisotropy direction in full Pantheon is close to the direction of the maximum $\mathrm{AL}_N$ and its angle difference is only $14.29^\circ$. In addition, the distribution of $\mathrm{AL}_N$ for the combination of Low-$z$ and SNLS subsamples is shown in Fig. \ref{fig:Number2}. The direction of the maximum $\mathrm{AL}_N$ is $(l,b)=({117.07^{\circ}},{8.39^{\circ}})$, which is $6.95^\circ$ away from the maximum anisotropy direction for the same combination. It may imply that the anisotropy level of the Pantheon sample significantly rely on the inhomogeneous distribution of SNe Ia in the sky.

\begin{figure}
	\begin{center}
		\includegraphics[width=8cm]{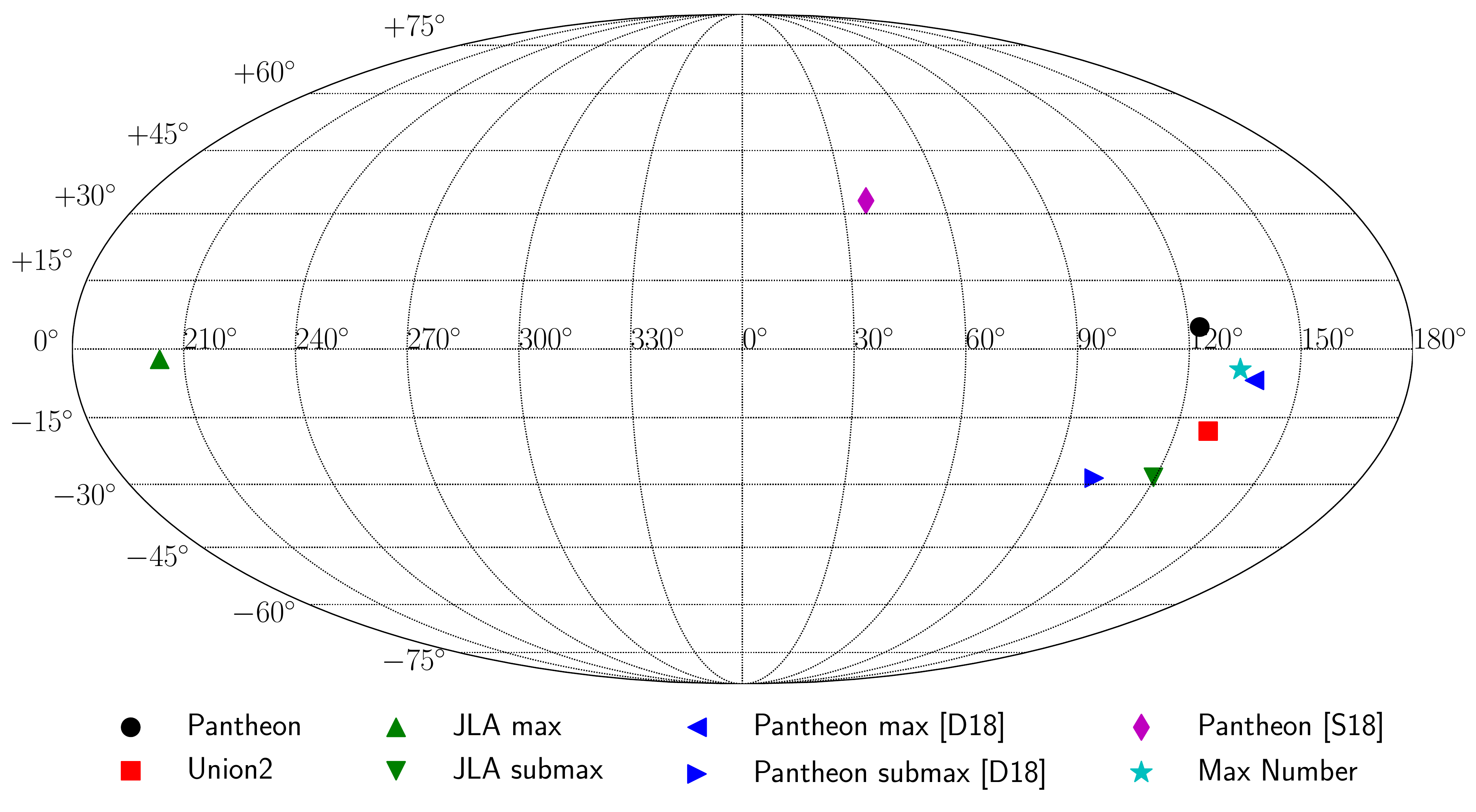}
		\caption{The HC preferred direction from the Pantheon sample and that from the Union2 sample \protect\citep{Antoniou:2010gw}, the JLA sample \protect\citep{Deng:2018yhb} and the Pantheon sample \protect\citep{Deng:2018jrp,Sun:2018cha}. The star marks the direction of the maximum number difference in the full Pantheon sample.}
		\label{fig:HCcompare}
	\end{center}
\end{figure}

At the end of this section, we make some comparisons between the HC preferred direction in the Pantheon sample with that derived from the Union2 sample \citep{Antoniou:2010gw}, the JLA sample \citep{Deng:2018yhb} and the Pantheon sample \citep{Deng:2018jrp,Sun:2018cha}. The HC preferred direction in each SNe Ia sample is shown in Fig. \ref{fig:HCcompare}, in the galactic coordinate system. As mentioned before, we find a similar anisotropy level map in the Pantheon sample with that in \citet{Deng:2018jrp} and \citet{Sun:2018cha}, but high resolution benefit for us to find the maximum anisotropy level. In \citet{Deng:2018jrp}, the maximum anisotropy level is $\mathrm{AL}_{max}=0.3088\pm0.0738$ and the corresponding direction $(l,b)=({138.08^{\circ}}^{+3.16^{\circ}}_{-16.90^{\circ}}, {-6.84^{\circ}}^{+13.55^{\circ}}_{-2.31^{\circ}})$, which is consistent with the preferred direction in this paper within $1\sigma$ error. However, the maximum anisotropy level in \citet{Sun:2018cha} is $\mathrm{AL}_{max}=0.136^{+0.009}_{-0.005}$, much less than 0.361 and the corresponding direction $(l,b)=({37^{\circ}}\pm{40^{\circ}},{33^{\circ}}\pm{16^{\circ}})$ totally deviates from the preferred direction in this paper, which could mean the 500 axes used in their work are not enough. The maximum anisotropy direction in the JLA sample \citep{Deng:2018yhb} is not consistent with that in this paper, but the submaximum anisotropy direction  $(l,b)=({119.47^{\circ}}^{+46.13^{\circ}}_{-23.39^{\circ}}, {-28.39^{\circ}}^{+17.01^{\circ}}_{-6.52^{\circ}})$ is about $33.35^\circ$ away from the preferred direction in this paper. Interestingly, in the Union2 sample \citep{Antoniou:2010gw}, the maximum anisotropy level is $\mathrm{AL}_{max}=0.43\pm0.06$ and the corresponding direction  $(l,b)=({129^{\circ}}^{+23^{\circ}}_{-3^{\circ}}, {-18^{\circ}}^{+10^{\circ}}_{-11^{\circ}})$ is $23.52^\circ$ away from the preferred direction in the Pantheon sample. In addition, we find these HC preferred directions from different SNe Ia samples are close to the direction of the maximum number difference in the full Pantheon. The consistency confirms the conclusion that the HC method strongly depends on the distribution of SNe Ia in the sky \citep{Chang:2014nca,Lin:2016jqp}.

\section{Dipole fitting method and result}\label{sec:DF}\noindent
The DF method was first employed to analyse the cosmic anisotropy using the Union2 SNe Ia sample \citep{Mariano:2012wx}. Then, it has been widely used for the spatial variation of fine structure \citep{Webb:2010hc,King:2012id} and MOND acceleration scale \citep{Chang:2018vxs}. We consider an expression of the theoretical apparent magnitude with monopole and dipole correction, namely
\begin{equation}
\widetilde{m}_{th}=m_{th}[1+A+B(\bm{\hat{n}} \cdot \bm{\hat{p}})],
\end{equation}
where $\bm{\hat{n}}$ is the dipole direction in the sky and $\bm{\hat{p}}$ is the unit vector pointing to the position of SNe Ia. $A$ and $B$ represent the monopole term and dipole magnitude, respectively. $m_{th}$ is the fiducial theoretical apparent magnitude predicted by the flat $\Lambda$CDM model given by Eq. \eqref{eq:mth}. In galactic coordinate system, the dipole direction can be represented by
\begin{equation}
\bm{\hat{n}}=\cos(l)\cos(b)\bm{\hat{i}}+\sin(l)\cos(b)\bm{\hat{j}}+\sin(b)\bm{\hat{k}},
\end{equation}
where $l$ and $b$ are the galactic longitude and latitude, respectively. $\bm{\hat{i}}$, $\bm{\hat{j}}$ and $\bm{\hat{k}}$ are unit vectors along the axis in the Cartesian coordinates system. Similarly, the position of the $i$th SNe Ia can be represented by
\begin{equation}
\bm{\hat{p}_i}=\cos(l_i)\cos(b_i)\bm{\hat{i}}+\sin(l_i)\cos(b_i)\bm{\hat{j}}+\sin(b_i)\bm{\hat{k}}.
\end{equation}

Two parameters $(\Omega_m,\,\mathcal{M})$ are used to predict the fiducial theoretical apparent magnitude. Four parameters $(A,~B,~l,~b)$ are used to describe the dipole anisotropy of the Universe. In total, there are six parameters used in the DF method. The best-fitting parameters can be derived by minimizing the corresponding $\chi^2$. We implement the Bayesian analysis to infer the best-fitting parameters and their $1\sigma$ error, by using the affine-invariant Markov chain Monte Carlo (MCMC) ensemble sampler in \textit{emcee}\footnote[1]{\url{https://emcee.readthedocs.io/en/stable/}} \citep{ForemanMackey:2012ig}, which is widely used in astrophysics and cosmology. In Bayesian analysis, the posterior distributions are determined by priors and likelihood functions, the latter is $\mathcal{L}\propto\exp{(-\chi^2/2})$. The \textit{emcee} could be regarded as a supplement of the CosmoMC \citep{Lewis:2002ah}. In this paper, one hundred random walkers are used to explore the entire parameter space. We run 500 iterations in the burn-in phase and another 2000 iterations in the production phase, which is sufficient for this analysis. We check that the acceptance fractions for all random walkers are in the range (0.2, 0.5). The flat prior on each parameter is used as follow: $\Omega_m\sim[0,1],~\mathcal{M}\sim[0,100],~A\sim[-1,1],~B\sim[0,1],~l\sim[-180,180^\circ],~b\sim[-90^\circ,90^\circ]$.

\begin{figure*}
	\centering
	\subfigure[Full Pantheon]{
		\includegraphics[width=6.75cm]{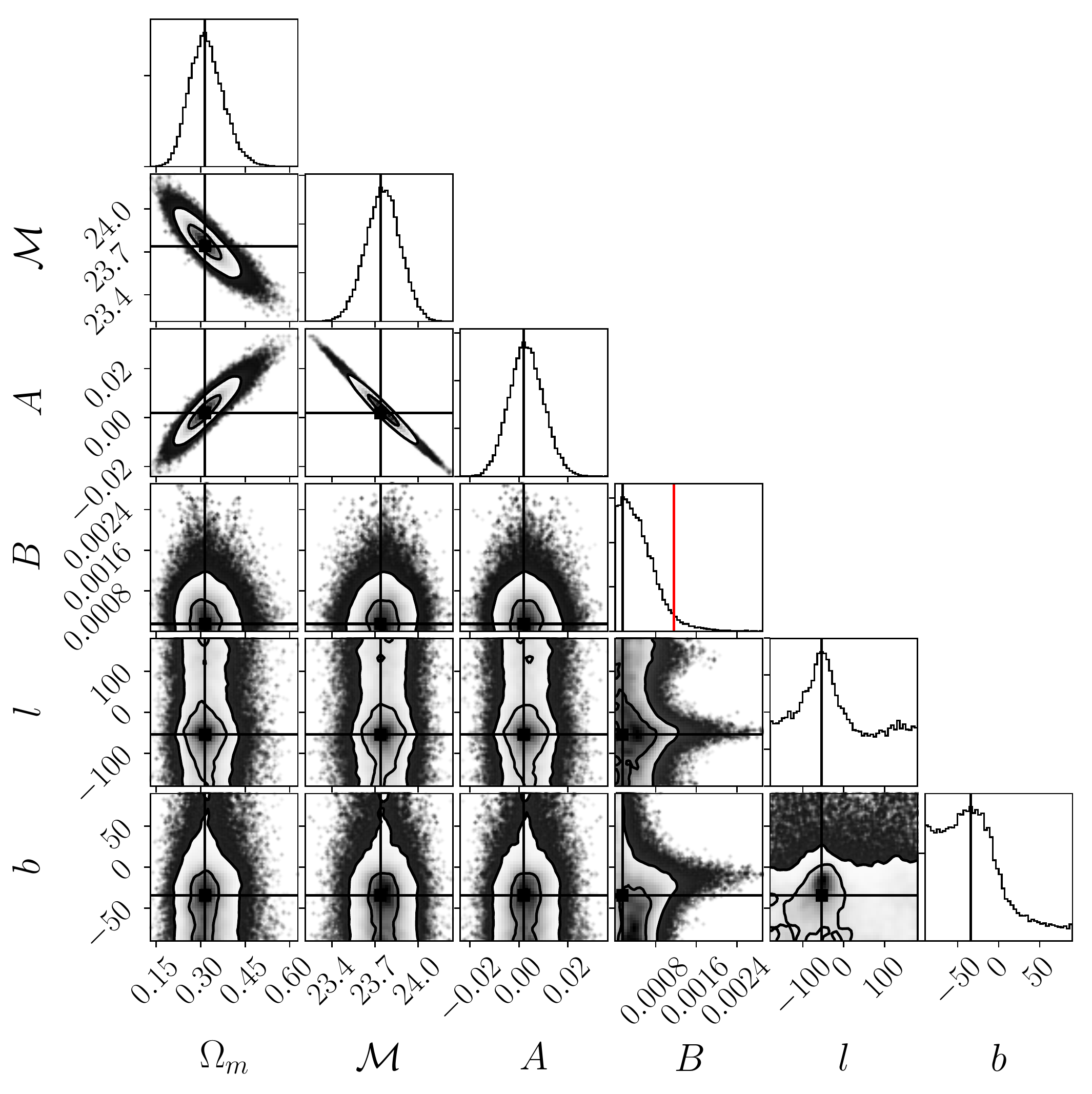} }
	\quad
	\subfigure[Pantheon without Low-$z$]{
		\includegraphics[width=6.75cm]{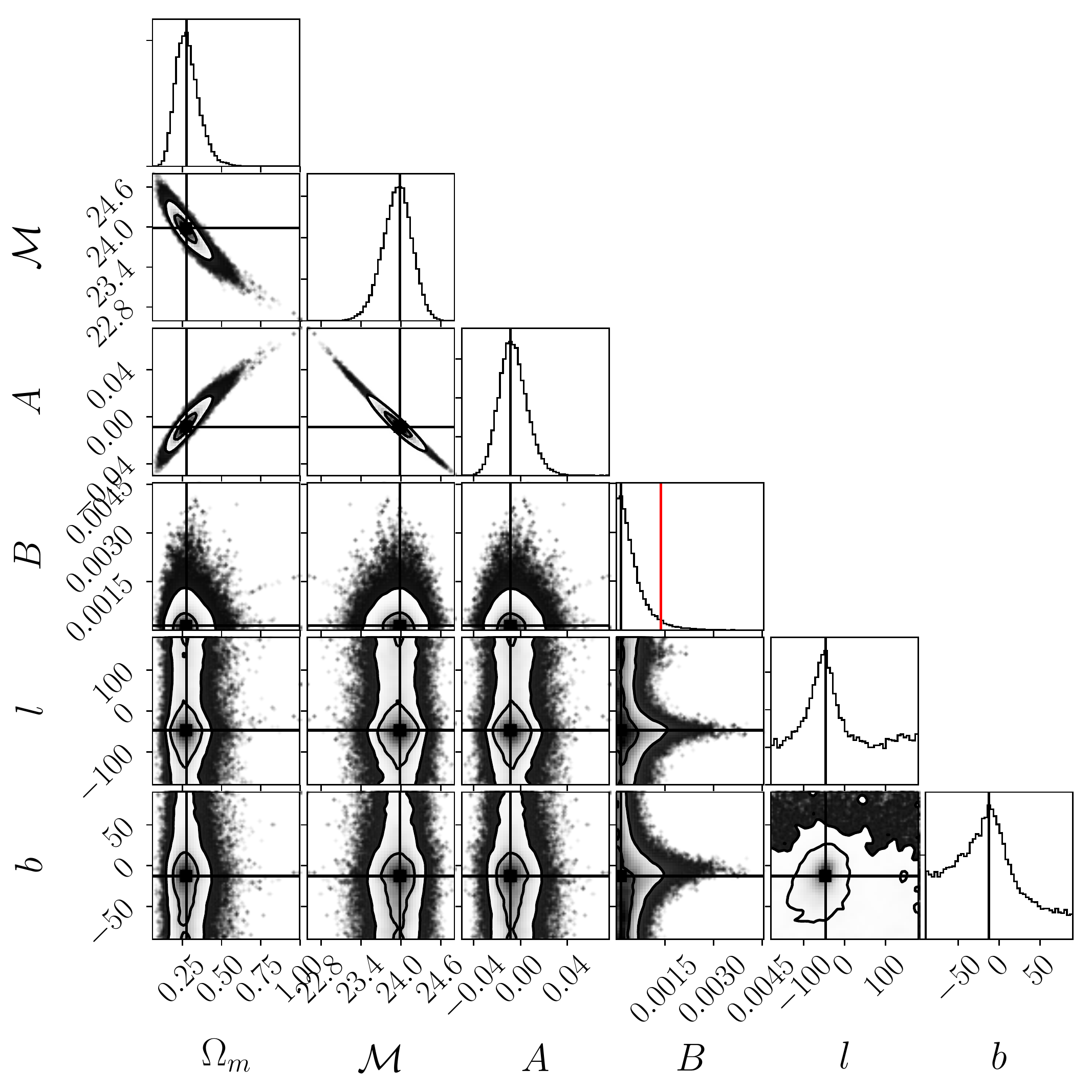} }
	\quad
	\subfigure[Pantheon without PS1]{
		\includegraphics[width=6.75cm]{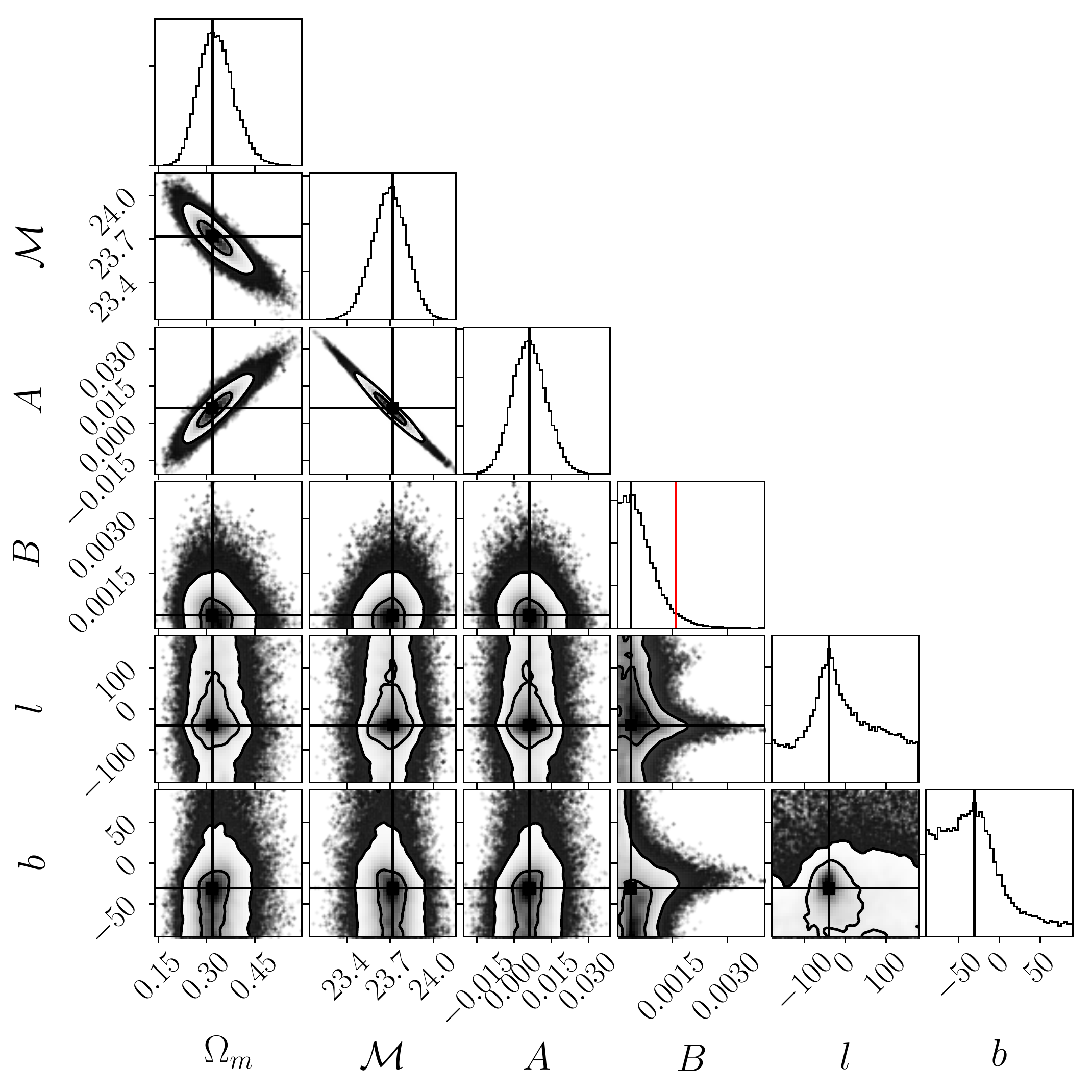} }
	\quad
	\subfigure[Pantheon without SDSS]{
		\includegraphics[width=6.75cm]{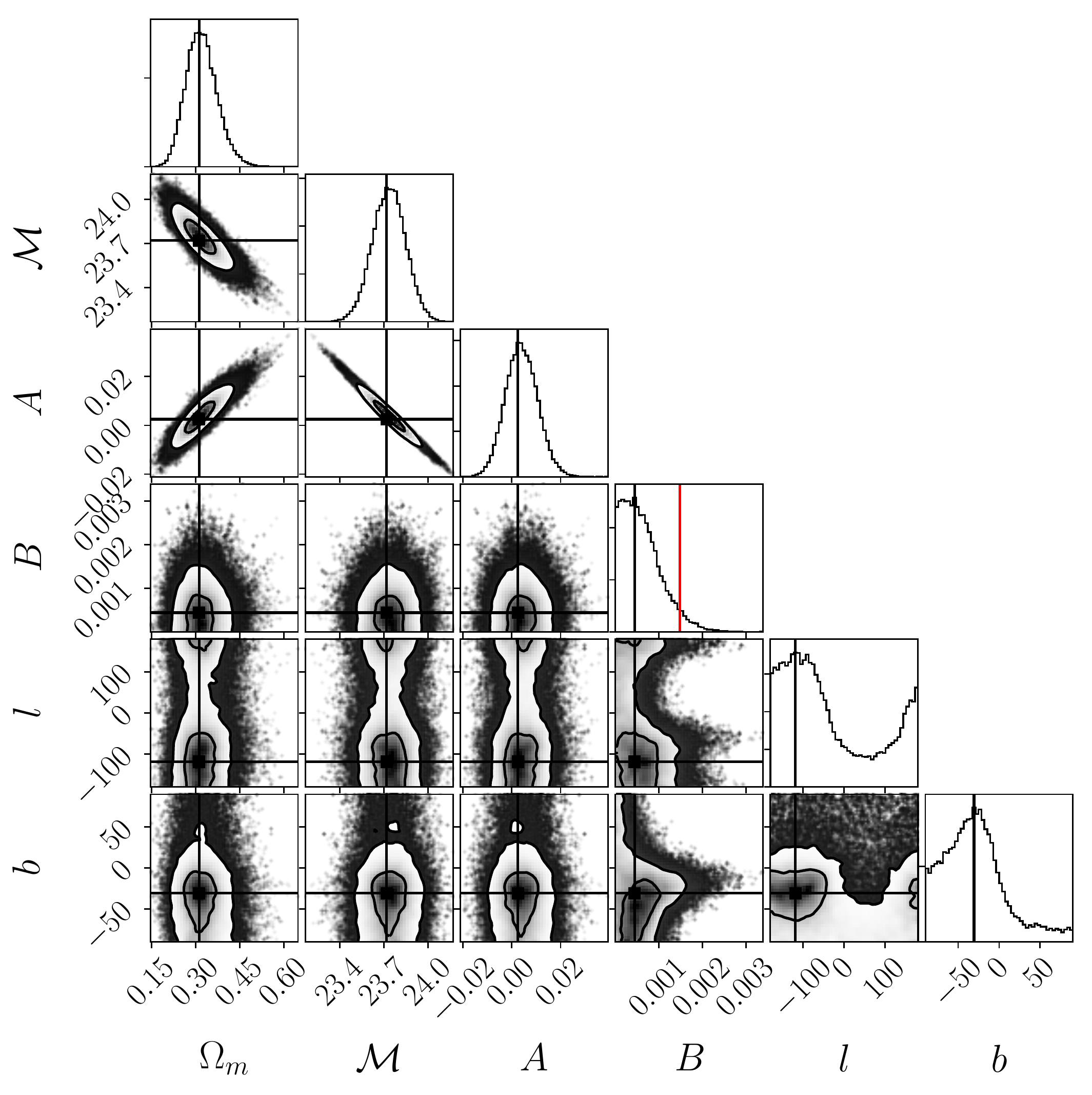} }
	\quad
	\subfigure[Pantheon without SNLS]{
		\includegraphics[width=6.75cm]{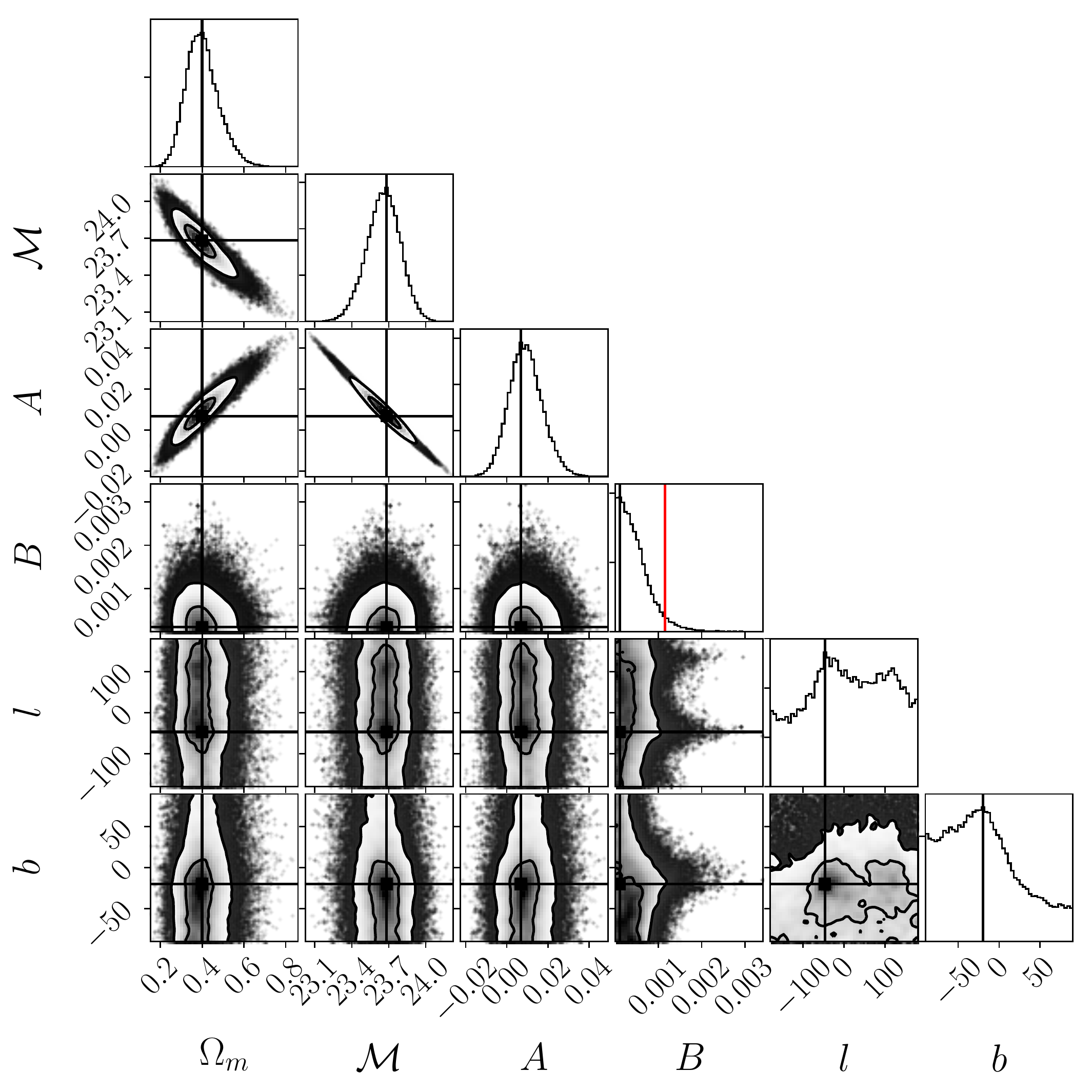} }
	\quad
	\subfigure[Pantheon without HST]{
		\includegraphics[width=6.75cm]{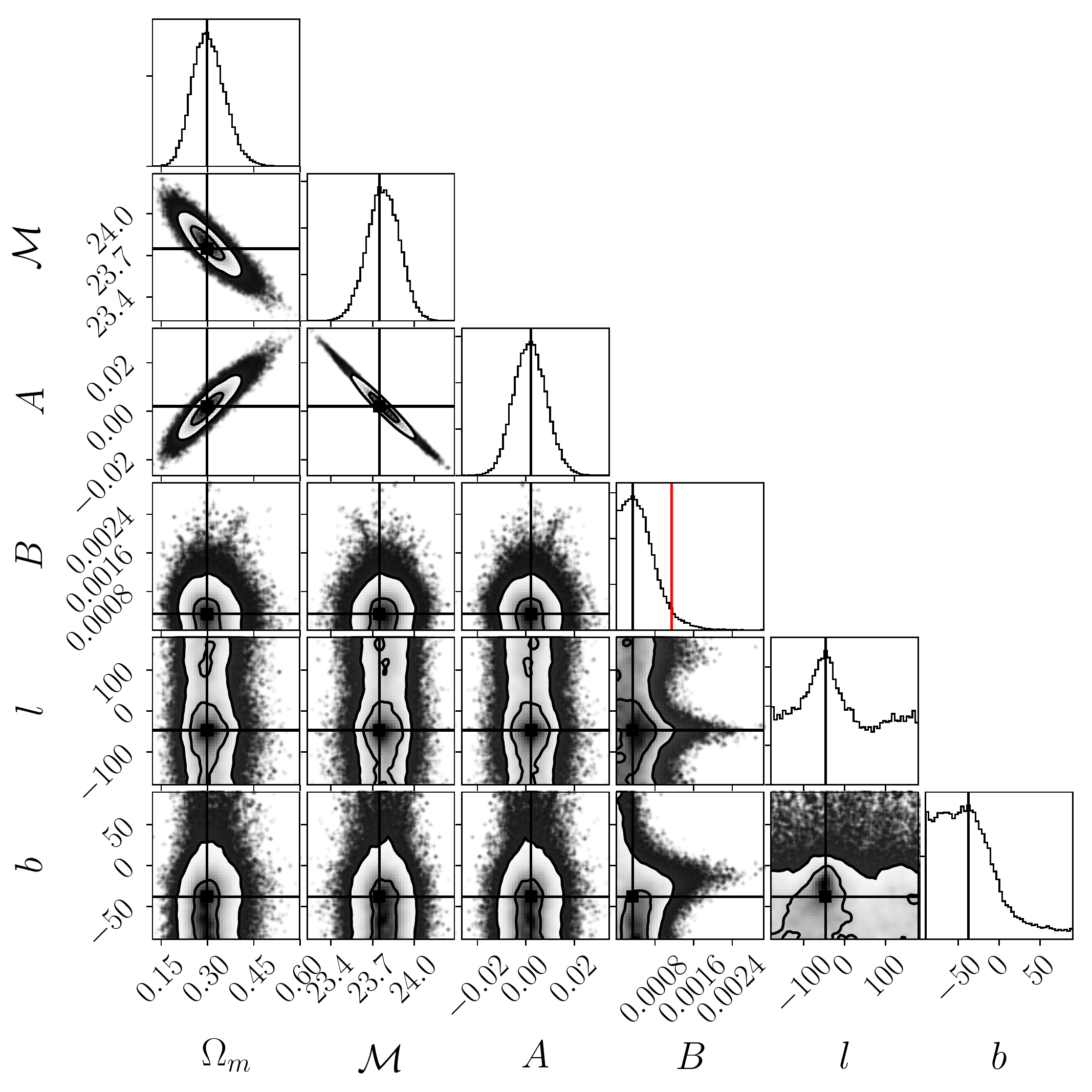} }
	\caption{The 1-dimensional and 2-dimensional marginalized posterior distributions by using the DF method to the full Pantheon sample and five combinations among Pantheon. The horizontal and vertical solid black lines mark the maximum of 1-dimensional marginalized posteriors. The red line indicates the $95\%$ CL upper limit of dipole magnitude $B$.}
	\label{fig:DF}
\end{figure*}

\begin{table*}
	\caption{The results of the DF method for the full Pantheon sample and five combinations among Pantheon. We show the maximum and its $68\%$ CL constraints on the model parameters $\Omega_m$, $\mathcal{M}$, $A$, $l$ and $b$, the $95\%$ CL upper limits of dipole magnitude $B$ in each combination.}
	\setlength{\tabcolsep}{2.85mm}{
		\begin{tabular}{lccccccc}
			\hline
			\multicolumn{1}{l}{Sample}                   &$N$   &$\Omega_m$                   &$\mathcal{M}$                 &$A[10^{-3}]$              &$B[10^{-3}]$ &$l[^{\circ}]$              &$b[^{\circ}]$                \\
			\hline
			\multicolumn{1}{l}{Full Pantheon}            &1048  &$ 0.315_{ -0.054}^{ +0.059}$ &$ 23.739_{ -0.102}^{ +0.140}$ &$ 1.91_{ -6.25}^{ +7.80}$ &$<1.16 $ &$ 306.00_{ -125.01}^{  +82.95}$ &$ -34.20_{ -54.93}^{ +16.82}$ \\
			\multicolumn{1}{l}{Pantheon without Low-$z$} &876   &$ 0.276_{ -0.075}^{ +0.068}$ &$ 23.985_{ -0.254}^{ +0.185}$ &$-8.64_{-10.54}^{+12.93}$ &$<1.37 $ &$ 313.20_{ -101.58}^{  +99.18}$ &$ -12.60_{ -58.98}^{ +28.55}$ \\
			\multicolumn{1}{l}{Pantheon without PS1}     &769   &$ 0.317_{ -0.047}^{ +0.062}$ &$ 23.720_{ -0.142}^{ +0.096}$ &$ 6.14_{ -7.37}^{ +6.44}$ &$<1.60 $ &$ 320.40_{  -57.75}^{ +135.85}$ &$ -30.61_{ -58.90}^{ +12.42}$ \\
			\multicolumn{1}{l}{Pantheon without SDSS}    &713   &$ 0.311_{ -0.049}^{ +0.056}$ &$ 23.719_{ -0.095}^{ +0.131}$ &$ 2.50_{ -5.24}^{ +7.89}$ &$<1.48 $ &$ 241.21_{  -84.08}^{  +76.31}$ &$ -30.60_{ -44.62}^{ +24.15}$ \\
			\multicolumn{1}{l}{Pantheon without SNLS}    &812   &$ 0.399_{ -0.082}^{ +0.070}$ &$ 23.682_{ -0.150}^{ +0.121}$ &$ 6.85_{ -6.65}^{ +9.46}$ &$<1.15 $ &$ 313.19_{  -27.58}^{ +188.66}$ &$ -19.80_{ -69.04}^{ +20.14}$ \\
			\multicolumn{1}{l}{Pantheon without HST}     &1022  &$ 0.298_{ -0.054}^{ +0.050}$ &$ 23.748_{ -0.092}^{ +0.142}$ &$ 2.08_{ -7.50}^{ +5.90}$ &$<1.15 $ &$ 313.21_{ -133.00}^{  +83.15}$ &$ -37.80_{ -52.11}^{ +12.76}$ \\
			\hline
	\end{tabular}}
	\label{table:DF}
\end{table*}

At first, we employ the DF method to investigate the cosmic anisotropy by using the full Pantheon sample. The marginalized posterior distributions of the DF method are shown in Fig. \ref{fig:DF}. We find the dipole anisotropy in the full Pantheon sample is very weak. The maximum of monopole term $A$ and the dipole magnitude $B$ are well consistent with zero within $1\sigma$. The monopole term is $A=(1.91_{ -6.25}^{ +7.80})\times10^{-3}$, and the dipole magnitude is less than $1.16\times10^{-3}$ at $95\%$ confidence level (CL). This result is consistent with that in \citet{Deng:2018jrp} at the same magnitude. Benefiting from the usage of \textit{emcee}, we obtain the detailed posterior distribution on the dipole direction. Fig. \ref{fig:DF} shows that the dipole direction is $(l,b)=({306.00^{\circ}}^{+82.95^{\circ}}_{-125.01^{\circ}}, {-34.20^{\circ}}^{+16.82^{\circ}}_{-54.93^{\circ}})$, here the galactic longitude is converted to positive value. The very large uncertainty of dipole direction also implies that the dipole anisotropy in the Pantheon sample is very weak. On the other hand, the maximum of $\Omega_m$ and $\mathcal{M}$ are almost the same as that without dipole anisotropy. The related results of the DF method are listed in Table \ref{table:DF}.

As same as the HC method,  we employ the DF method to those five combinations among Pantheon. The marginalized posterior distribution for each combination is shown in Fig. \ref{fig:DF} and all related results are summarized in Table \ref{table:DF}. We find the dipole anisotropy in each combination is also very weak. The dipole direction is consistent with that in the full Pantheon sample, except one combination has large deviation. For the Pantheon without SDSS subsample, the dipole direction is $(l,b)=({241.21^{\circ}}^{ +76.31^{\circ}}_{-84.08^{\circ}}, {-30.60^{\circ}}^{+24.15^{\circ}}_{-44.62^{\circ}})$, about $53.89^\circ$ away from the dipole direction in full Pantheon. Note that there are 335 SNe Ia in the SDSS subsample clustering in a narrow strip which corresponds to the equator of the equatorial coordinate system. Coincidentally, the axial direction of equator with galactic coordinates $(l,b)=(302.93^\circ,-27.13^\circ)$ is very close to the dipole direction in the full Pantheon sample and its angle difference is only $7.55^\circ$. Therefore, the SDSS subsample could be the decisive part that impacts the dipole anisotropy in the full Pantheon sample. Similar conclusions have been found in the Union2 sample \citep{Jimenez:2014jma}.

\begin{figure}
	\begin{center}
		\includegraphics[width=8cm]{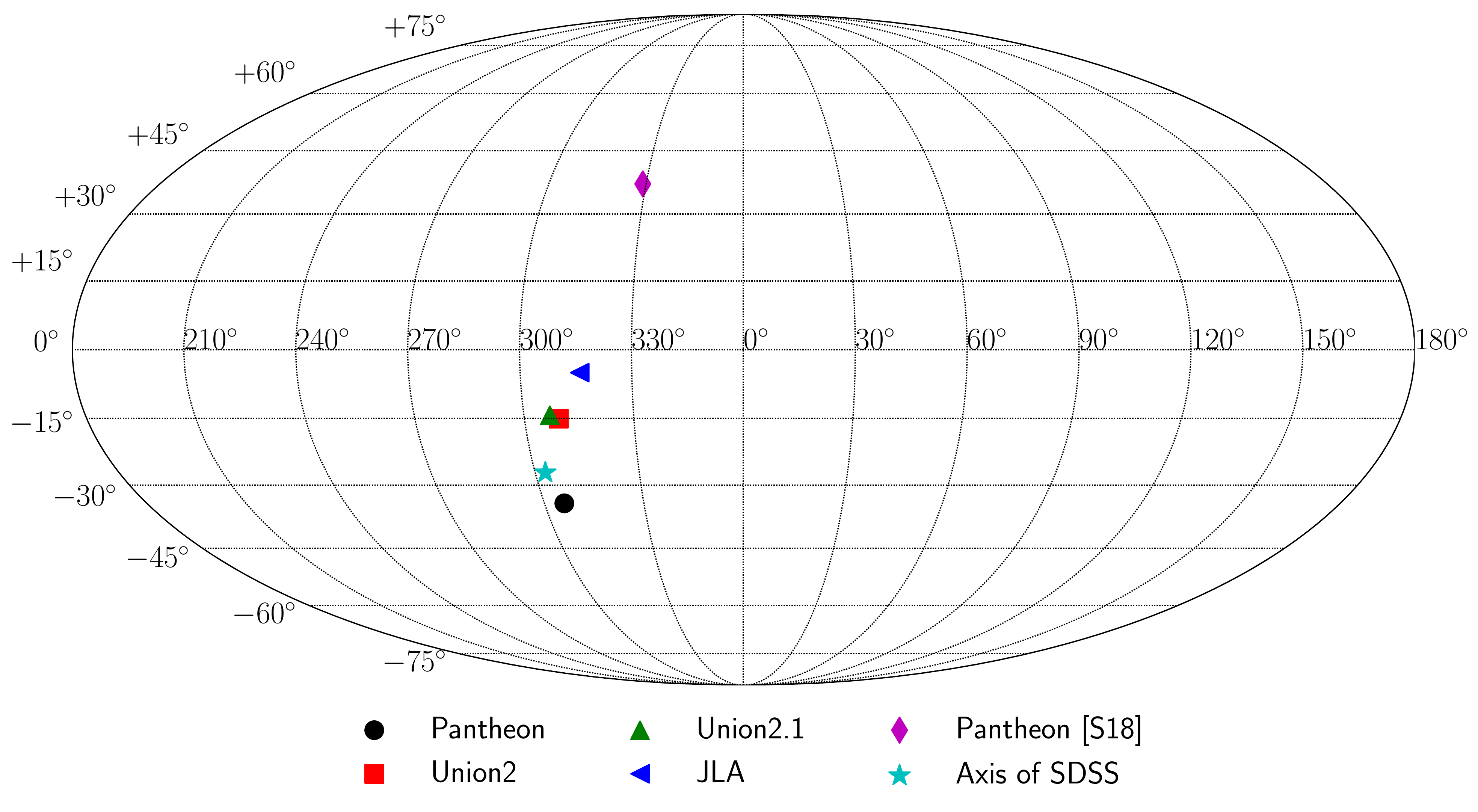}
		\caption{The dipole direction from the Pantheon sample and that from the Union2 sample \protect\citep{Mariano:2012wx}, the Union2.1 sample \citep{Yang:2013gea}, the JLA sample \protect\citep{Lin:2015rza} and the Pantheon sample \protect\citep{Sun:2018cha}. The star marks the axial direction to the plane of SDSS subsample.}
		\label{fig:DFcompare}
	\end{center}
\end{figure}

At the end, we make some comparisons between the dipole direction in the Pantheon sample with that derived from the Union2 sample \citep{Mariano:2012wx}, the Union2.1 sample \citep{Yang:2013gea}, the JLA sample \citep{Lin:2015rza} and the Pantheon sample \citep{Sun:2018cha}. The dipole direction in each SNe Ia sample are shown in Fig. \ref{fig:DFcompare}, in the galactic coordinate system. As we can seen, the dipole direction in \citet{Sun:2018cha} is $(l,b)=({329^{\circ}}^{ +101^{\circ}}_{-28^{\circ}}, {37^{\circ}}^{+52^{\circ}}_{-21^{\circ}})$, totally deviates from all other dipole directions. Even though the dipole direction in JLA sample \citep{Lin:2015rza} is  $(l,b)=({316^{\circ}}^{ +107^{\circ}}_{-110^{\circ}}, {-5^{\circ}}^{+41^{\circ}}_{-60^{\circ}})$, about $30.64^\circ$ away from the dipole directions in the full Pantheon sample. However, we find the dipole direction in Pantheon and JLA sample are close to the dipole direction in the Union2 or Union2.1 sample \citep{Mariano:2012wx,Yang:2013gea}. All of these dipole directions are close to each other hinting that there could exist an underlying relation. The star in Fig. \ref{fig:DFcompare} indicates the axial direction to the plane of  SDSS subsample,  which is well close to these dipole directions, especially for the dipole direction in the Pantheon sample. Since the large deviation for the Pantheon without SDSS subsample, it may imply that the dipole anisotropy in the Pantheon sample is mostly determined by the distribution of the SDSS subsample, and other subsamples only have small impact.

\section{Conclusions}\label{sec:conclusion}\noindent
The cosmological principle has been extensively tested by SNe Ia and the cosmic anisotropy is preferred by most observations. However, the distribution of SNe Ia is very inhomogeneous in the sky, which could bring significant impact on the cosmic anisotropy.
In this paper, we employed the hemisphere comparison (HC) method and the dipole fitting (DF) method to investigate the cosmic anisotropy in the full Pantheon sample as well as five combinations among Pantheon. The combinations are obtained by excluding individual subsamples from the Pantheon sample in turn.

For the HC method, we found the anisotropy level map of the full Pantheon sample is similar to the results in \citet{Deng:2018jrp} and \citet{Sun:2018cha}. Benefiting from the usage of HEALPix, we found the maximum anisotropy level is $\mathrm{AL}_{max}=0.361\pm0.070$ and corresponding direction is $(l,b)=({123.05^{\circ}}^{+11.25^{\circ}}_{-4.22^{\circ}}, {4.78^{\circ}}^{+1.80^{\circ}}_{-8.36^{\circ}})$. In this direction, the Universe has the minimum accelerating expansion while the maximum one is in the opposite direction. As a robust check, we found the maximum anisotropy level from the full Pantheon sample  is hardly reproduced by simulated isotropic data set and its statistical significance is about $2.1\sigma$.  The numbers of SNe Ia in ``up'' and ``down'' hemispheres are 883 and 165, respectively. The inhomogeneous distribution of Pantheon SNe Ia in the sky could have significant impact on the anisotropy level in HC method. According to the anisotropy level map in Fig. \ref{fig:HC}, we found that the HC preferred direction in the Pantheon without Low-$z$ or SNLS subsample has large deviation from the HC preferred direction in the full Pantheon sample. We also employed the HC method to the combination of Low-$z$ and SNLS subsamples. We found a consistent preferred direction with that in the full Pantheon sample. All of these results show the Low-$z$ and SNLS subsamples have decisive impact on the HC result of the full Pantheon sample. In addition, we found the direction of the maximum number difference is close to the maximum anisotropy direction for the full Pantheon sample and the combination of low-$z$ and SNLS subsample. It may imply that the anisotropy level of the Pantheon sample signicantly rely on the inhomogeneous distribution of SNe Ia in the sky. At the end, we made some comparisons with the HC preferred direction in the Union2 sample \citep{Antoniou:2010gw}, the JLA sample \citep{Deng:2018yhb} and the Pantheon sample \citep{Deng:2018jrp,Sun:2018cha}. Interestingly, we found the preferred direction in the Pantheon sample is close to that in the Union2 sample, and its angle difference is only $23.52^\circ$. The consistency confirms the conclusion that the HC method strongly depends on the distribution of SNe Ia in the sky \citep{Chang:2014nca,Lin:2016jqp}.

For the DF method, we implemented the Bayesian analysis by using the MCMC method to infer the best-fitting parameters and their $1\sigma$ error. According to the marginalized posterior distributions in Fig. \ref{fig:DF}, we found the dipole anisotropy in the Pantheon sample is very weak. The monopole term and the dipole magnitude are well consistent with zero within $1\sigma$. The monopole term is $A=(1.91_{ -6.25}^{ +7.80})\times10^{-3}$, and the dipole magnitude is less than $1.16\times10^{-3}$ at $95\%$ CL. However, the dipole direction is well inferred by MCMC method and it points towards $(l,b)=({306.00^{\circ}}^{+82.95^{\circ}}_{-125.01^{\circ}}, {-34.20^{\circ}}^{+16.82^{\circ}}_{-54.93^{\circ}})$. The DF method was also employed to each combination among Pantheon. We found the dipole direction in the Pantheon without SDSS subsample has a large deviation, about $53.89^\circ$ away from the dipole direction in the full Pantheon sample. In addition, the axial direction to the plane of SDSS subsample is very close to the dipole direction in the full Pantheon and its angle difference is only $7.55^\circ$. It implies that the SDSS subsample could be the decisive part that impacts the dipole anisotropy in the full Pantheon sample. We also found the dipole direction in the Union2 sample is close to that in the Pantheon sample, and the SDSS subsample could also be the decisive part to the dipole anisotropy in Union2 sample.

All of these results imply that the cosmic anisotropy found in Pantheon sample significantly rely on the inhomogeneous distribution of SNe Ia in the sky. More homogeneous distribution of SNe Ia is necessary to search for a convincing cosmic anisotropy.

\section*{Acknowledgements}\noindent
We thank the anonymous referee for valuable suggestions and comments. We appreciate Dr. Y. Sang for useful discussions. This work has been funded by the National Natural Science Foundation of China under Grant Nos. 11675182 and 11690022.

\label{lastpage}

\end{document}